%% file: main.tex
\documentclass[pdflatex,sn-vancouver-num]{sn-jnl}% Vancouver Numbered Reference Style
\usepackage[T1]{fontenc} % get rid of missing 'Font shape `OMS/cmss/m/n' undefined'-warning of the template

\usepackage{anyfontsize} % get rid of the missing rsfs front sizes

\usepackage{bookmark} % replaces standard hyperref bookmark handling and natively resolves level-jump warnings

\usepackage{mathtools, cuted} % enables one line math equations in double column environments

\usepackage{graphicx}%
\usepackage{silence}% must precede caption: filters the warning declared just below
\usepackage[font=small]{caption}% load before subcaption, which would otherwise pull caption in optionless
\usepackage{subcaption}%
\usepackage{multirow}%
\usepackage{amsmath,amssymb,amsfonts}%
\usepackage{mathrsfs}%
\usepackage[title]{appendix}%
\usepackage{xcolor}%
\usepackage{textcomp}%
\usepackage{manyfoot}%
\usepackage{booktabs}%
\usepackage{algorithm}%
\usepackage{algorithmicx}%
\usepackage{algpseudocode}%
\usepackage{listings}%
\usepackage{lipsum}
\usepackage{url}

\usepackage{makecell}

\DeclareCaptionLabelSeparator{bar}{ $|$ }
\usepackage{tikz}
\usetikzlibrary{patterns}

\definecolor{indianred}{HTML}{cd5c5c}
\definecolor{color_Qs}{HTML}{440154}
\definecolor{color_streaks}{HTML}{31688e}
\definecolor{color_chong}{HTML}{35b779}
\definecolor{tab_red}{HTML}{d62728}
\definecolor{tab_blue}{HTML}{1f77b4}
\definecolor{tab_green}{HTML}{2ca02c}

\theoremstyle{thmstyleone}%
\theoremstyle{thmstyletwo}%

\theoremstyle{thmstylethree}%

\unnumbered

\begin{document}

\title[Article Title]{Unveiling wing turbulence dynamics through explainable deep learning}


\author*[1]{\fnm{Samuel} \sur{Molina-Casino}}\email{samuel.molina\_casino@onera.fr}

\author[2]{\fnm{Andrés} \sur{Cremades}}

\author[2]{\fnm{Sergio} \sur{Hoyas}}

\author[1]{\fnm{José I.} \sur{Cardesa}}

\author[1]{\fnm{François} \sur{Chedevergne}}

\author*[3,4]{\fnm{Ricardo} \sur{Vinuesa}}\email{rvinuesa@umich.edu}

\affil[1]{\orgdiv{Multi-Physics Department for Energetics}, \orgname{ONERA}, \orgaddress{\city{Toulouse}, \postcode{F-31055}, \country{France}}}

\affil[2]{\orgdiv{Instituto Universitario de Matemática Pura y Aplicada}, \orgname{Universitat Politècnica de València}, \orgaddress{\city{Valencia}, \postcode{46022}, \country{Spain}}}

\affil[3]{\orgdiv{Department of Aerospace Engineering}, \orgname{University of Michigan}, \orgaddress{\city{Ann Arbor}, \postcode{MI 48109}, \country{United States}}}

\affil[4]{\orgdiv{FLOW, Engineering Mechanics}, \orgname{KTH Royal Institute of Technology}, \orgaddress{\city{Stockholm}, \postcode{SE-100 44}, \country{Sweden}}}

\abstract{
The three-dimensional organization of coherent structures in the turbulent flow over aircraft wings remains elusive, limiting our ability to reduce fuel consumption. Here, we use explainable artificial intelligence to characterize these structures by predicted flow evolution rather than classical predefined kinematic criteria. After training a deep neural network to predict the short-term evolution of the flow, we identify the highest-relevance flow regions using Shapley-value attribution methods. We have found that as the flow decelerates towards the trailing edge, predictive importance shifts from near-wall low-speed structures to three-dimensional pairs of high- and low-speed fluid regions. Matching no single classical structure family, these pairs progressively dominate the dynamically relevant flow. Most stand spanwise side-by-side, enclosing a near-vertical momentum interface with the strongest velocity jump. Their geometry remains invariant in viscous units while their volume expands when approaching separation. Our results reveal a predictive organization that traditional paradigms overlook, opening new ways to control turbulent flows.
%Turbulence over aircraft wings drives drag, impacting fuel consumption. Yet the three-dimensional organisation of its coherent structures remains elusive. We define these structures by their contribution to predicting flow evolution, rather than predefined kinematic criteria, through explainable deep learning. A deep neural network predicts short-term evolution of a turbulent boundary layer developing on a wing section's suction side; Shapley-value attribution then identifies highest-relevance velocity-field regions. As flow decelerates towards the trailing edge, predictive importance shifts from near-wall low-speed structures to three-dimensional pairs of high- and low-speed fluid. Matching no single classical structure family, these pairs progressively dominate the dynamically relevant flow. Most stand spanwise side-by-side, enclosing a near-vertical momentum interface with the strongest velocity jump among all configurations. Their geometry remains invariant in viscous units while their volume expands approaching separation. Our results provide a new way to understand wing turbulence dynamics, revealing spatial organisations overlooked by traditional paradigms.
}




\maketitle

%%==================================%%
%%==================================%%
% \section{Introduction}
% \label{sec:introduction}
\phantomsection
\pdfbookmark[1]{Introduction}{sec:introduction}

% Ricardo (v3):
Extracting physical mechanisms from high-dimensional simulations remains a central challenge in computational science. Numerical solvers can resolve complex multiscale dynamics and deep-learning-based methods can predict their short-term evolution, but neither of them can identify the spatial distributions most informative about that evolution. Scientific interpretation therefore still relies heavily on predefined observables, which can encode assumptions inherited from simpler systems and overlook mechanisms that emerge in more complex regimes.

Turbulent boundary layers over aircraft wings provide a societally relevant test of this problem because they generate aerodynamic drag and associated fuel consumption~\cite{iata_global_2024, hoyas_role_2026}. On the suction side, the flow decelerates under an adverse pressure gradient (APG), thickening the boundary layer, shifting the turbulent activity away from the wall and eventually producing flow separation~\cite{anderson_fundamentals_2017}. The discrepancy with respect to canonical wall-bounded turbulence and zero-pressure-gradient (ZPG) flows is quantified by the Rotta--Clauser pressure-gradient parameter $\beta=\delta^*/\tau_w(\mathrm{d}P_e/\mathrm{d}x_t)$ (where $\delta^*$ is the displacement thickness, $\tau_w$ the wall-shear stress and $\mathrm{d}P_e/\mathrm{d}x_t$ the wall-parallel pressure gradient at the boundary-layer edge)~\cite{pope_turbulent_2000}. As $\beta$ increases, the wall-normal transport strengthens and an increasing fraction of the turbulent energy is carried by outer-layer motions~\cite{spalart_experimental_1993, skare_turbulent_1994, monty_parametric_2011, vinuesa_turbulent_2018, tanarro_effect_2020}. However, the commonly reported turbulence statistics do not reveal which instantaneous three-dimensional events are responsible for these complex phenomena.

Much of our current understanding of wall turbulence is built around coherent structures~\cite{jimenez_coherent_2018}: spatially organised motions associated with a substantial fraction of momentum and energy transport. In canonical wall-bounded flows, kinematic criteria applied to instantaneous velocity fields have revealed streamwise streaks~\cite{kline_structure_1967}, intense Reynolds-stress events~\cite{lozano-duran_three-dimensional_2012}, vortex clusters~\cite{chong_general_1990} and larger outer-layer motions~\cite{jimenez_coherent_2018, meinhart_existence_1995, silva_uniform_2016, silva_interfaces_2017}. Applying the same criteria to APG boundary layers has shown that known energetic events become less prevalent near the wall and increasingly detached as the pressure gradient strengthens~\cite{maciel_coherent_2017, atzori_coherent_2020, atzori_control_2022}. However, the definitions of these structures rely on prior hypotheses based on the dynamics of ZPG boundary layers. When applied under APG, they can quantify how the classically defined structures respond to deceleration, but they cannot determine whether different types of motions become dynamically dominant as the flow departs from the canonical regime. This illustrates a broader limitation of hypothesis-driven feature extraction in high-dimensional scientific data: it can test structures specified in advance, but it cannot discover spatio-temporal organisations that become relevant only after the system leaves the regime in which those definitions were developed.

% Samuel (v1):
% Much of our recent progress in understanding turbulent flows rests on the concept of coherent structures~\cite{jimenez_coherent_2018}: connected regions that keep their identity over space and time and carry a large part of the momentum and energy transfer. They are identified with kinematic criteria applied to the instantaneous velocity field, which detect streamwise streaks~\cite{kline_structure_1967}, intense Reynolds-stress events~\cite{lozano-duran_three-dimensional_2012} or vortex clusters~\cite{chong_general_1990}. In wall-bounded flows without a pressure gradient, these criteria uncovered the current picture: a near-wall cycle of streaks and quasi-streamwise vortices, with larger-scale motions organising the outer layer~\cite{jimenez_coherent_2018}. The latter is also described as uniform-momentum zones separated by thin internal shear layers~\cite{meinhart_existence_1995,silva_uniform_2016,silva_interfaces_2017}, a picture assembled almost entirely from streamwise--wall-normal planes. The same criteria have been applied under APG, showing that intense events become less numerous near the wall and more often detached from it~\cite{maciel_coherent_2017,atzori_coherent_2020,atzori_control_2022}. These studies show how known structures respond to APG, but not whether they remain the relevant ones: motions that become important under strong deceleration but match no canonical criterion may go unnoticed.

Explainable deep learning offers a way to identify the most informative flow regions without prescribing what they should look like. In this work, we train a convolutional neural network to predict the short-term evolution of the flow from an instantaneous velocity field, so the model must rely on the regions that most influence that evolution. SHAP (SHapley Additive exPlanations)~\cite{shapley_value_1953, lundberg_unified_2017}, a game-theoretic method, then assigns to every network input its average contribution to the prediction. The resulting three-dimensional attribution field therefore ranks the different flow regions by how informative they are for the predicted future state. Thresholding this field yields connected regions which we call importance structures, defined by their predictive relevance rather than by a predefined kinematic criterion.

Previous applications of this framework have focused on canonical turbulent channel flow. There, importance structures overlap only partially with streaks and intense Reynolds-stress events, and the regions with the largest predictive importance are not necessarily those contributing most strongly to Reynolds shear stress~\cite{cremades_identifying_2024, cremades_classically_2025}. They are recovered from two-dimensional slices \cite{molina-casino_inferring_2025} and from experimental fields \cite{cremades_identifying_2024}, and they are physically meaningful rather than merely statistical: a deep-reinforcement-learning agent trained to target them reduces drag more than one aimed at classical structures \cite{beneitez_improving_2025}. Building on those studies, we extend the framework to a boundary layer under APG as it approaches separation.

The data come from a high-resolution large-eddy simulation (LES) of the turbulent boundary layer over a NACA~4412 wing section at $Re_c=2\times10^5$ and $5^\circ$ angle of attack, computed with the spectral-element solver \textsc{Nek5000} \cite{fischer_nek5000_2008, vinuesa_turbulent_2018}. Along the suction side the Clauser parameter grows continuously from near-ZPG conditions $\beta \approx 0.1$ to near separation $\beta \approx 10$ within one simulation (Fig.~\ref{fig1}b). The scale sets the computational problem: 10{,}000 instantaneous velocity fields on an interpolated grid with 44 million points and a surrogate with 3.3 million trainable parameters. Previous work on this configuration established the identification procedure and its relationship with classical structure families~\cite{molina-casino_uncovering_2026}. Here we address the broader physical question of how the flow organisation carrying predictive information changes as the APG strengthens. We classify the importance structures by their velocity content and determine how their relative relevance, spatial arrangement and internal dynamics evolve along the wing, thereby testing whether the structures that become dominant under strong deceleration can still be described within the classical picture of wall turbulence. This opens the way to data-driven sensing, flow-control and modelling strategies that target the structures most relevant to the flow dynamics, rather than those selected a priori by classical criteria.

% Here we go beyond that comparison, classifying the structures by their velocity content and tracking how their organisation, dynamics and relative relevance change along the wing as the pressure gradient intensifies.

% Samuel (v2):
% We consider a high-resolution large-eddy simulation (LES) of the turbulent boundary layer over a NACA~4412 wing section at $Re_c=2\times10^5$ and $5^\circ$ angle of attack \cite{vinuesa_turbulent_2018}. Along the suction side, the Clauser parameter grows continuously from near-ZPG conditions $\beta \approx 0.1$ to near separation $\beta \approx 10$ in a single simulation (Fig.~\ref{fig1}b). Previous work established the identification procedure and its relationship with classical structure families~\cite{molina-casino_uncovering_2026}. Here, we go beyond that comparison by classifying the importance structures according to their velocity content and tracking how their organisation, dynamics and relative relevance change along the wing as the pressure gradient intensifies. 

%%==================================%%
%%==================================%%
\section{Results}
\label{sec:results}

Every result below follows the same analysis framework (Fig.~\ref{fig1}a, Methods). A U-Net predicts the near-future velocity field and gradient-SHAP attributes that prediction back to the input grid points. Thresholding the resulting importance field identifies $4.3\times10^6$ structures across 6{,}000 snapshots. Throughout, $x$, $y$ and $z$ denote the streamwise, wall-normal and spanwise directions on the suction side, and $u$, $v$ and $w$ the corresponding velocity fluctuations. Positions are often normalised by the chord $c$, and the superscript $+$ marks quantities scaled in viscous units (Methods).

%%==================================%%
%\subsection{Importance concentrates in high/low-speed pairs, not in intense events}
\subsection{Predictive importance shifts from low-speed structures to high/low-speed pairs}
\label{sec:census}

\begin{figure*}[p!]
    \centering
    \includegraphics[width=0.95\linewidth]{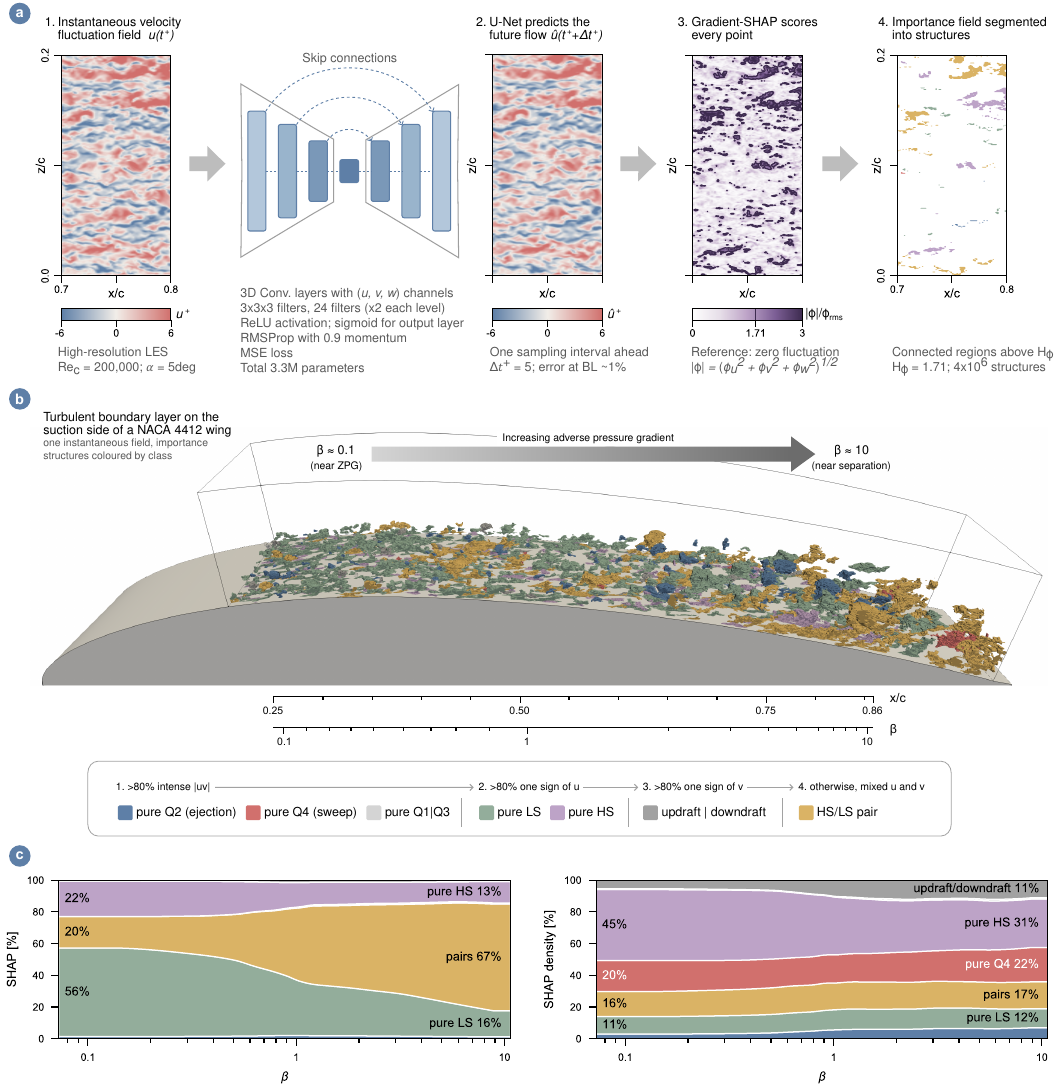}
    \caption{\textbf{Detection, classification and census of the importance structures in the adverse-pressure-gradient turbulent boundary layer.}
    \textbf{a}, A U-Net trained on high-resolution large-eddy simulation fields of a NACA~4412 wing section at $Re_c=2\times10^5$ and $5^\circ$ angle of attack predicts the three velocity-fluctuation components one sampling interval ahead, to within ${\sim}1\%$ error inside the boundary layer. Gradient-SHAP then scores every grid point against a zero-fluctuation reference, giving the importance vector $\mathbf{\Phi}$. The connected regions where $|\mathbf{\Phi}|$ exceeds $H_\Phi$ times its local root-mean-square (rms) define the importance structures. Contours are illustrative, taken on a wall-parallel slice with axes normalised by the chord $c$.
    \textbf{b}, Importance structures on the suction side for one instantaneous field, coloured by class. Classes follow the tree below: motions with intense Reynolds shear stress give pure Q2 (ejection, blue), pure Q4 (sweep, red) and pure Q1$|$Q3 (light grey); single-signed $u$ gives pure LS (green) and pure HS (purple); single-signed $v$ gives updraft$|$downdraft (dark grey); everything else is an HS/LS pair (gold). The wire-frame axes outline the analysis domain, $0.25<x/c<0.86$, over which the Clauser parameter $\beta$ rises from ${\approx}0.1$ to ${\approx}10$.
    \textbf{c}, Share of the total importance held by each class versus $\beta$ (left) and the same share per unit structure volume, the importance density (right), as stacked areas summing to 100\%. Colours as in \textbf{b}.}
    \label{fig1}
\end{figure*}

The importance field identifies the regions of the input that contribute most strongly to the prediction, but not the type of motion those regions represent. We therefore design a taxonomy that assigns every importance structure a class from the velocity fluctuations over its grid points (the classification tree of Fig.~\ref{fig1}, Methods). Pure Q structures are dominated by intense Reynolds shear stress; pure LS (low-speed) or pure HS (high-speed) by uniformly slow or fast flow; updrafts or downdrafts by a dominant vertical velocity. The HS/LS pairs are the remainder: fast and slow fluid bound into one structure with neither a dominant speed nor a coherent vertical motion. The slow ($u<0$) and fast ($u>0$) regions of a pair are its lobes, which makes the pairs the only double-lobed class. The Q classes take their names from the quadrants of classical wall turbulence~\cite{lozano-duran_three-dimensional_2012}. Ejection-like motions (Q2) lift slow fluid away from the wall while sweep-like motions (Q4) push fast fluid towards it; together they are the sweep-ejection motions, down-gradient in that they act to reduce the mean velocity gradient, transporting momentum. Pure Q1 and pure Q3 act the other way, and are the outward-inward, or counter-gradient, motions. These class names describe the velocity content of structures identified independently, through their predictive importance.

The classical picture of wall turbulence places particular emphasis on regions of intense momentum transfer, and several widely used criteria are designed specifically to isolate such events \cite{wallace_wall_1972, lozano-duran_three-dimensional_2012}. To test that expectation, every class is given its share of five budgets: structure count, volume, importance score, Reynolds shear stress and turbulent kinetic energy (Methods; Supplementary Note~3). Together, these budgets provide a census of the structures identified by the model. Fig.~\ref{fig1}c follows the importance budget and its density along the chord. Near the leading edge, where $\beta\approx0.1$, pure LS structures hold 56\% of the importance: the fluid that matters there moves uniformly slowly, the signature of the streak-dominated near-wall cycle of canonical wall turbulence~\cite{jimenez_coherent_2018}. By $\beta\approx10$ that share has fallen to 16\%, while the HS/LS pairs have risen from 20\% to 67\% of the importance and to a matching two thirds of the flagged volume, overtaking the pure LS and pure HS structures at $\beta\approx1$. The outward-inward quadrants, pure Q1 and pure Q3, are negligible in every budget, so the whole reorganisation happens within the sweep-ejection motions that produce absolute Reynolds shear stress. Pooled over the chord the ranking is the same: pairs are one object in six yet hold about half of every budget, while pure Q2 and pure Q4 structures take 2\% of the importance between them.

Total importance and importance per unit volume give opposite rankings, which sharpens the same reading. Dividing each budget by class volume inverts the order: pure HS and pure Q4 (sweep) structures are the densest in importance and in Reynolds shear stress, while pure Q2 (ejection) structures are the most dilute and pure LS structures sit close to the population average (Fig.~\ref{fig1}c, right; Supplementary Table~3). The inversion holds at every $\beta$, so point for point the most informative fluid is fast even though slow fluid fills more of the flagged volume. Sweeps are already known to dominate wall-friction generation~\cite{hoyas_deep-learning-based_2025} and to gain weight near the wall under APG~\cite{houra_effects_2000}; the density ranking extends that role from momentum transfer to predictive relevance.

The census therefore reveals a systematic reorganisation of predictive importance as the APG strengthens. Low-speed structures dominate under near-canonical conditions, but that organisation disappears as the flow decelerates. Importance shifts towards HS/LS pairs, which bind high- and low-speed fluid within one connected region and become the dominant class well before the boundary layer approaches separation.

%%==================================%%
\subsection{Beyond classical structures}
\label{sec:relabelling}

The classical families of wall turbulence are the natural reference for what the importance field selects. A structure holding fast and slow fluid together could be a low-speed streak beside its high-speed neighbour, or an ejection paired with a sweep. Measuring the two sets against each other establishes whether the importance structures are a relabelling of those families. Three such families are therefore detected on the same snapshots, each with its standard criterion and a threshold fixed by the same percolation analysis (Methods): intense Q events~\cite{lozano-duran_three-dimensional_2012}, high- and low-speed streamwise streaks~\cite{kline_structure_1967}, and vortex clusters~\cite{chong_general_1990}. Two point-wise measures make the comparison: the composition, the fraction of the importance volume belonging to a given family, and the capture, the fraction of that family flagged as important.

\begin{figure*}[tp!]
    \centering
    \includegraphics[width=\linewidth]{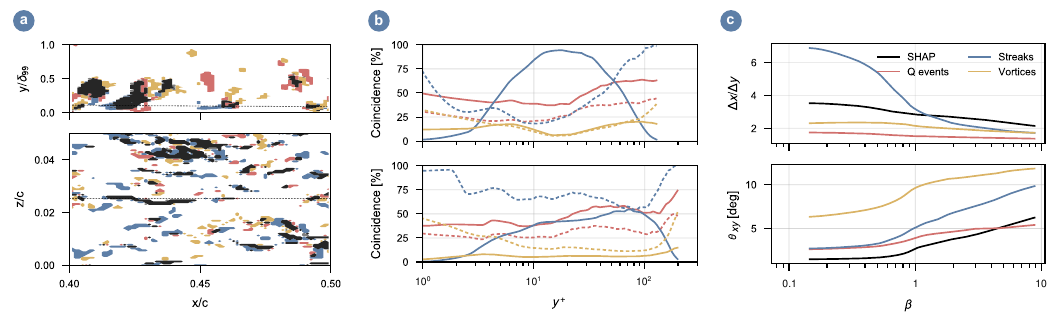}
    \caption{\textbf{Spatial overlap and geometry of the importance structures relative to the three classical families of coherent structures.}
    \textbf{a}, Two orthogonal slices through one instantaneous field, showing where the importance structures fall relative to the three classical families detected from the same field with their own percolation-derived thresholds (Methods): Q events (red), streamwise streaks (blue) and vortex clusters (gold). Importance structures are grey and their intersection with any family is black; grey is barely visible because almost all of the importance volume in this field overlaps at least one family. Top, streamwise--wall-normal plane, with the wall distance normalised by the boundary-layer thickness $\delta_{99}$; bottom, wall-parallel plane. The dashed line in each panel marks the position of the other slice.
    \textbf{b}, Voxel-wise coincidence between the importance structures and each family versus wall distance $y^+$, at a weak-APG station ($x/c=0.3$, top) and a strong-APG station ($x/c=0.85$, bottom); colours as in \textbf{a}. Solid, composition of the importance volume $(X\cap S)/S$, the fraction of the importance-structure voxel set $S$ belonging to family $X$; dashed, capture rate $(X\cap S)/X$, the fraction of family $X$ flagged as important. The dashed curves rising towards 100\% at the outermost heights are a geometric effect of the vanishing family volumes there.
    \textbf{c}, Median streamwise-to-wall-normal aspect ratio $\Delta x/\Delta y$ (top) and median inclination $\theta_{xy}$ in the streamwise--wall-normal plane (bottom) versus $\beta$, for the importance structures (black, all classes pooled) and the three families (colours as in \textbf{a}).}
    \label{fig2}
\end{figure*}

A single instantaneous field already shows how tightly the importance structures and the classical families are interleaved: most importance structures overlap at least one classical family, yet no family covers them (Fig.~\ref{fig2}a). The overlap statistics (Fig.~\ref{fig2}b) reveal a departure from the classical picture as the APG increases. At the weak-APG station ($x/c=0.3$, $\beta\approx0.3$) the composition is a streak monopoly, 93\% of the important fluid in the buffer layer being streak fluid, with Q events contributing at all heights and vortices a small minority. By the strong-APG station ($x/c=0.85$, $\beta\approx10$) the streak peak is gone and the composition spreads across families and across heights. The capture rates sharpen this reading: the near-wall streaks that survive are almost entirely flagged as important, while the Q-event capture stays moderate at both stations (the streak criterion alone uses friction velocity rather than a local rms, so its curves at the strongest stations carry that conditioning; Methods). Vortex capture remains the lowest at any APG, extending the channel-flow results~\cite{cremades_classically_2025} to strong APG flows. 

Shape does not separate the families either (Fig.~\ref{fig2}c, Supplementary Note~7). At weak APG each family carries its own geometric signature, with streaks strongly elongated, importance structures intermediate, and vortices and Q events compact. Deceleration erases those signatures, and by the strongest stations all four have converged to nearly the same compact proportions.
%APG evolution uncovers yet another trend (Fig.~\ref{fig2}c, Supplementary Note~7). At weak APG each family has its own signature, with streaks strongly elongated (median length-to-height ratio $\Delta x/\Delta y=6.9$), importance structures intermediate (3.5), and vortices (2.3) and Q events (1.8) compact. Deceleration erases these signatures. At strong APG the four families converge to nearly the same compact proportions, so what distinguishes them can no longer be their shape but must lie in what they contain and how they are arranged.

No combination of the three families accounts for the importance set. At strong APG about a quarter of the important volume lies outside all of them, and that volume is moderate-amplitude fluid close to the local mean velocity, which every intensity criterion discards by construction (Supplementary Figs.~2 and 17). Under strong deceleration the classical families are therefore no longer sufficient to describe the regions that matter to the prediction, and what sets those regions apart is no longer their shape but what they contain and how they are arranged.
%No combination of the three families accounts for the importance set, and at strong APG about a quarter of the important volume lies outside all of them. This volume is moderate-amplitude fluid close to the local mean velocity that every intensity criterion discards (Supplementary Figs.~2 and 17).

%%==================================%%
% \subsection{The anatomy of the high/low-speed pairs}
\subsection{High/low-speed pairs enclose momentum interfaces with a few recurrent geometries}
\label{sec:pairs}

\begin{figure*}[tp!]
    \centering
    \includegraphics[width=0.95\linewidth]{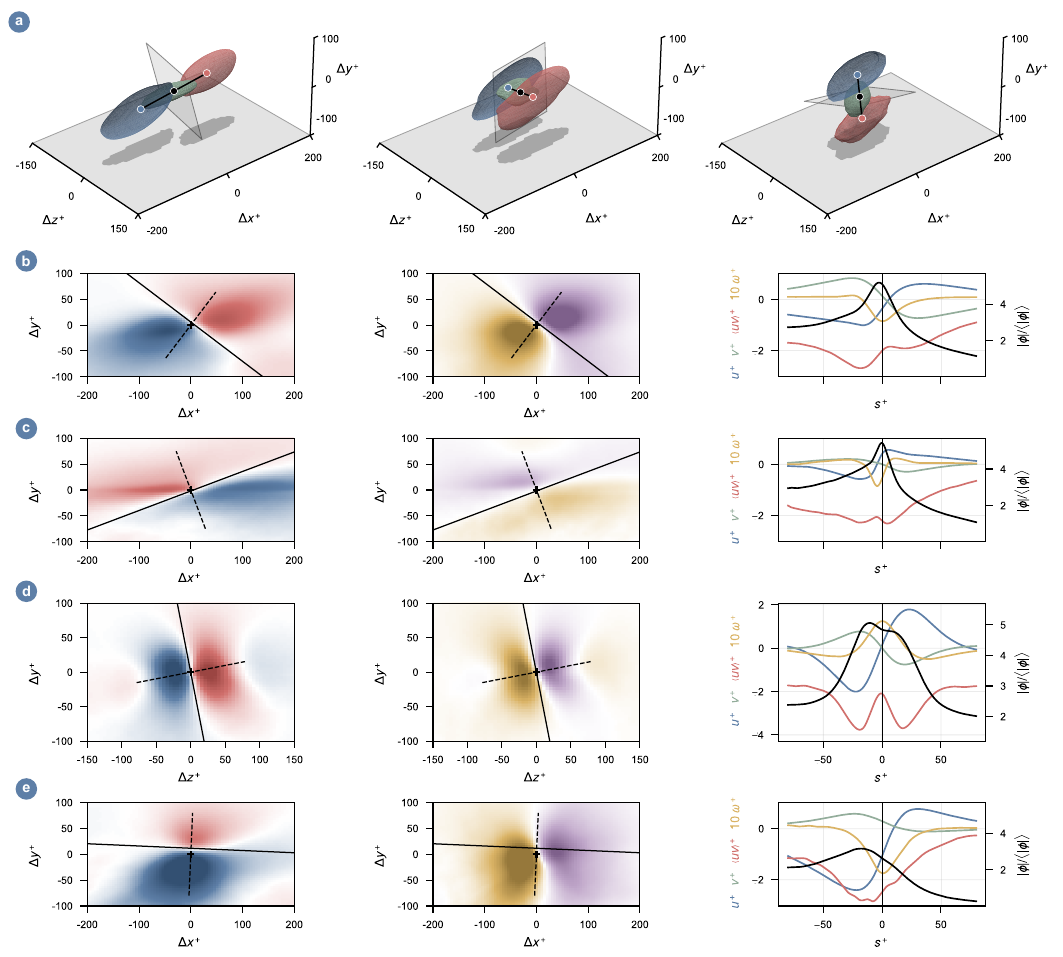}
    \caption{\textbf{Arrangement and interface structure of the HS/LS pairs.}
    \textbf{a}, Three-dimensional conditional averages of the three arrangement axes: streamwise, spanwise and vertical, from left to right, each centred on the midpoint between the lobe centroids and expressed in wall-unit offsets. The spanwise panel shows the orientation with the fast lobe at positive $\Delta z^+$; the opposite is equally common and quantitatively similar. Blue and red isosurfaces enclose 50\% of the low- and high-speed composite peaks; green, the top 0.2\% of the importance. Markers: low-speed centroid (blue), high-speed centroid (red), structure centroid (black). The translucent grey sheet is the fitted $u=0$ interface.
    \textbf{b}--\textbf{e}, Planar composites and interface-normal profiles for (\textbf{b}) streamwise pairs with the fast lobe downstream (the inclined shear layer), (\textbf{c}) streamwise pairs with the fast lobe upstream (the burst front), (\textbf{d}) spanwise pairs and (\textbf{e}) vertical stacks with the fast lobe on top. The spanwise composite is shown in the cross-flow plane ($\Delta z^+$, $\Delta y^+$); the others in the streamwise--wall-normal plane ($\Delta x^+$, $\Delta y^+$); the cross marks the composite centre. Left column, streamwise velocity $u^+$ (blue to red, $\pm1.35$, red positive); middle, wall-normal velocity $v^+$ (purple to gold, $\pm0.60$, gold positive); black lines, fitted $u^+=0$ interface (solid) and its normal (dashed). Right column, profiles along the interface normal $s^+$ (interface at $s^+=0$, low-speed side to the left): left axis showing $u^+$ (blue), $v^+$ (green), $(uv)^+$ (red), $10\,\omega^+$ (gold); and $|\mathbf{\Phi}|/\langle|\mathbf{\Phi}|\rangle$ (black) on the right axis. Each arrangement is averaged over its full population.}
    \label{fig3}
\end{figure*}

The offset from the slow lobe centroid to the fast one describes how a pair is arranged. Its direction sorts the pairs onto three axes, streamwise, spanwise and vertical (Fig.~\ref{fig3}a), each with two orientations, hence six arrangements (Methods). Pooled by axis, spanwise pairs hold 55\% of the pair volume, streamwise pairs 38\% and vertical stacks 5\%: the most common way the model binds fast fluid to slow fluid is side-by-side across the span (Fig.~\ref{fig3}a, middle). The same ranking holds at every pressure-gradient strength (Supplementary Table~4 and Fig.~4). By count the spanwise and streamwise pairs are equally numerous, so the spanwise ones are individually larger, with compact and steady lobe separations in wall units (Supplementary Figs.~7 and 8).

Conditional averages (composites) over each arrangement make the pairs' internal flow quantitative (Methods; full atlas in Supplementary Note~5). They return the same object every time: two counter-moving bodies of fluid on either side of a thin surface where the streamwise fluctuation changes sign, with the importance concentrated on that surface. The arrangements differ in how that interface is oriented and how sharp it is.

Each arrangement then resolves into a recognisable flow feature already described in the literature. The streamwise pair with the fast lobe downstream matches the well-documented inclined internal shear layer (Fig.~\ref{fig3}b): slow fluid upstream and below rises while fast fluid downstream and above sinks, converging on a thin interface inclined at $37^\circ$ and leaning upstream as it rises. The 10--90\% velocity jump across the interface is $\Delta u^+=1.6$ over about 26 wall units, matching published pipe and boundary-layer values \cite{silva_interfaces_2017, gul_internal_2020} and recovered without prescribing any shear layer. The reversed arrangement, with fast fluid upstream, corresponds to the near-wall burst front~\cite{jeong_coherent_1997} in which a fast patch overruns a slow streak; it is shallower, less than half as thick, and the weakest in vertical motion (Fig.~\ref{fig3}c). The vertical stack resembles the step between two uniform-momentum zones caught as a single object, with the jump-to-thickness proportions reported for zone edges~\cite{silva_interfaces_2017}; its vertical velocity is a streamwise up-and-downwash couple rather than a lobe-wise exchange, consistent with a quasi-static zone boundary rather than an active momentum pump (Fig.~\ref{fig3}e).

The majority arrangement is the one the classical literature has least often seen. The spanwise pair encloses a near-vertical momentum interface carried by streamwise vorticity, the largest of the three arrangement axes by volume and the strongest (Fig.~\ref{fig3}d). Two counter-moving streams stand side-by-side across the span, the slow lobe rising and the fast lobe sinking, separated by a $u=0$ surface leaning only about $11^\circ$ off the wall-normal. A streamwise roll produces exactly this arrangement through the lift-up effect, carrying slow near-wall fluid up on one flank and fast outer fluid down on the other~\cite{brandt_lift-up_2014}, so the sharpest spanwise change in streamwise velocity falls between the lobes. Its jump reaches $\Delta u^+=3.8$ across about the same 26 wall units, the steepest of the four. The two orientations are mirror images and track each other along the ramp (Supplementary Fig.~10 and Table~5).

Across all arrangements the importance concentrates on that $u=0$ surface, biased a few wall units into the low-speed side, with a peak-to-background contrast of about a factor of two (Fig.~\ref{fig3}, right column). Because the model is never told about zones or interfaces, this is evidence, independent of any structural hypothesis, that the fluid most informative for predicting the short-term evolution of the flow sits on momentum interfaces.

\begin{figure*}[tp!]
    \centering
    \includegraphics[width=\linewidth]{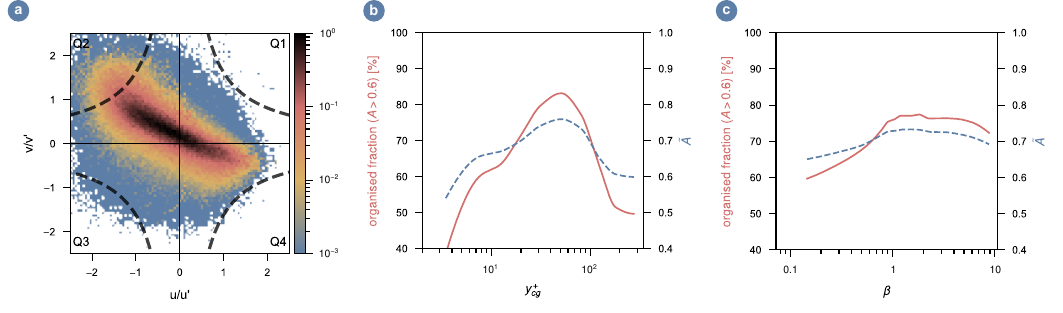}
    \caption{\textbf{Fluid content and vertical organisation of the HS/LS pairs.}
    \textbf{a}, Volume-weighted joint distribution of the velocity fluctuations at the grid points inside HS/LS pairs, in the quadrant plane of $u/u'$ against $v/v'$. Dashed lines indicate the frontier defined by the $H_{uv}=1.62$ criterion on intense Q events (Methods); quadrants are labelled in the corners. Colour, joint density on a logarithmic scale from $10^{-3}$ to $1$, least to most populated (blue to black).
    \textbf{b}, Fraction of pairs that are strongly organised, taken as a sweep-ejection alignment $A>0.6$ (red solid, left axis), and median alignment $\tilde{A}$, which involves no threshold (blue dashed, right axis), versus the wall distance $y^+$ of the pair centroid. $A$ is the volume imbalance between rising and sinking fluid inside a pair. The $A>0.6$ cut is the alignment reached when four fifths of the deciding lobe moves the same way (Methods).
    \textbf{c}, The same two quantities versus $\beta$.}
    \label{fig4}
\end{figure*}

Importance and Reynolds shear stress peak in different places. The fluid inside a pair does run along the sweep-ejection diagonal, from the ejection quadrant through the origin into the sweep quadrant, but at moderate amplitude, so most of its volume lies inside the hyperbolic hole that intensity criteria discard (Fig.~\ref{fig4}a). The composites separate the two fields directly: the importance rises to a single maximum on the interface, where the stress has a local minimum, while the stress peaks in the lobe interiors on either side (Fig.~\ref{fig3}, right column, black and red curves). The same dissociation holds between the two lobes of a single object, where the low-speed lobe carries about four times the Reynolds shear stress of its high-speed counterpart at weak APG, closing to about $1.5$ times by the end of the ramp, yet the model weighs the two almost equally at every station (Supplementary Note~6). Made within one object, that comparison is unaffected by the collapse of friction velocity $u_\tau$ along the chord. The pairs are therefore interfaces rather than intense events, and what makes them relevant is the interface itself, not the momentum flux on either side of it. Hence, the channel-flow result that importance and stress do not coincide~\cite{cremades_identifying_2024, cremades_classically_2025} holds for APG flows and within single objects.

%%==================================%%
\subsection{Interface volume grows with APG while its geometry stays fixed}
\label{sec:apg_response}

The census showed the pairs taking a rising share of the importance and of the flagged volume, but not how they come to hold it. Deceleration could sharpen each interface, making it better organised, steeper or stronger, or it could leave the object as it is and simply give it more of the flow. Separating the two means following the arrangements along the ramp and asking what about them actually changes.

Whether the interfaces are arranged so as to promote momentum transfer is measured object by object. Each pair is scored by a sweep-ejection alignment $A$, the volume imbalance between rising and sinking fluid inside its lobes (Methods). Sorted by that alignment, the pairs form a continuous distribution rather than splitting into active and passive populations, so they differ in organisation by degree and not by kind (Supplementary Note~2). What sets the degree of organisation is height rather than pressure gradient. Both the median alignment and the fraction above the $A>0.6$ cut rise through the buffer layer, peak near $y^+=40$--$100$ and fall away towards the outer flow (Fig.~\ref{fig4}b), so the pairs transfer momentum most effectively at intermediate wall distances. The median involves no threshold, so that trend belongs to the population and not to where the cut is placed. Against $\beta$, by contrast, both curves rise only until $\beta\approx1$ and change little thereafter (Fig.~\ref{fig4}c), while the census shows the pair volume growing to the end of the ramp. Past that point the pairs claim more volume without becoming any better organised.

\begin{figure*}[tp!]
    \centering
    \includegraphics[width=\linewidth]{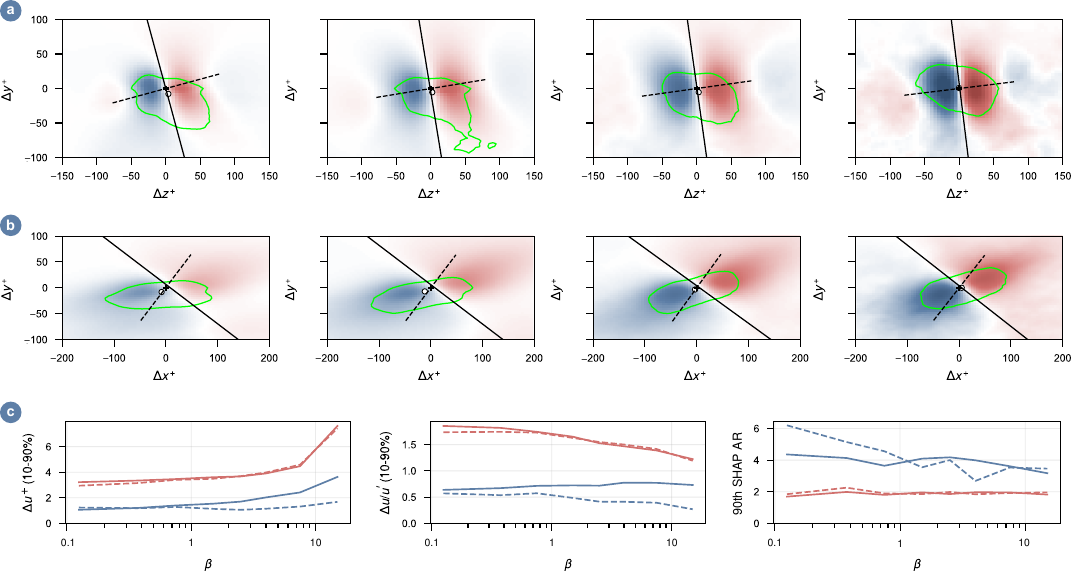}
    \caption{\textbf{Interface composites, velocity jump and importance geometry as functions of the Clauser parameter $\beta$.}
    \textbf{a}, Conditional average of the streamwise velocity in viscous units for the spanwise pairs, rebuilt within four bands of $\beta$ (left to right: $\beta<1$, $1$--$5$, $5$--$10$, $\ge10$) and shown in the cross-flow plane; colour scale and choice of orientation as in Fig.~\ref{fig3}. Green contour, 90th percentile of the importance within the composite, its centroid marked by the white circle; black lines, fitted $u=0$ interface (solid) and its normal (dashed); cross, composite centre.
    \textbf{b}, The same for the streamwise pairs, in the streamwise--wall-normal plane.
    \textbf{c}, Left, 10--90\% velocity jump across the interface in friction units, $\Delta u^+$; middle, the same jump normalised by the local turbulence intensity $u'$, the root-mean-square streamwise fluctuation; right, aspect ratio of the 90th-percentile importance region. All three versus $\beta$. Red, spanwise pairs (solid, fast lobe to the right; dashed, fast lobe to the left); blue, streamwise pairs (solid, fast lobe downstream, the inclined shear layer; dashed, fast lobe upstream, the burst front).}
    \label{fig5}
\end{figure*}

Geometry answers the same way. Rebuilt band by band in $\beta$, the interface composites change little (Fig.~\ref{fig5}a,b): the inclined shear layer holds its angle to within a few degrees across the four $\beta$ bands and its thickness at 25--30 wall units, while the spanwise interface keeps its thickness and steepens slightly towards the wall-normal (Supplementary Note~5). The region the model attends to is just as fixed: the aspect ratio of the 90th-percentile importance contour stays near 2 for the spanwise pairs and between 3 and 6 for the streamwise ones, declining somewhat towards separation (Fig.~\ref{fig5}c, right).

Strength is the one property that does change. In friction units the velocity jump grows by a factor of 2--3 along the ramp (Fig.~\ref{fig5}c, left), as expected because friction scaling must fail as $u_\tau\rightarrow0$. Measured against the local turbulence intensity $u'$, the jumps stay bounded and of order unity but do not collapse (Fig.~\ref{fig5}c, middle). Both scalings are conditioned by the attribution baseline, which carries the local fluctuation as a multiplicative factor (Methods), so these ratios are reported as measurements, not as scalings. The inclined shear layer stays near $\Delta u\approx0.6\,u'$ across the whole analysis domain, the spanwise interface drifts from about $1.9$ to $1.2$, keeping roughly twice the contrast of the inclined layer, and the burst front falls from about $0.6$ to about $0.3$, so how deceleration strengthens an interface depends on which way it faces.

The takeover of Fig.~\ref{fig1}c is therefore expansion and not concentration. The importance contrast of a single interface weakens along the ramp and every class loses absolute importance density, so deceleration dilutes the importance field rather than sharpening it. What grows is the volume the interfaces hold: the volume of pure LS fluid is flat across two orders of magnitude of $\beta$, while the pair volume rises by more than an order of magnitude (Supplementary Note~3). Deceleration does not dismantle the near-wall cycle. The momentum contrasts that surround it come to hold most of the flagged fluid, a gain in volume that far outpaces any change in their shape or their organisation.

%%==================================%%
%%==================================%%
\section{Discussion}
\label{sec:discussion}

At negligible pressure gradient a boundary layer keeps its predictive information in the fluid moving uniformly slowly near the wall. As deceleration strengthens, that information migrates into HS/LS pairs of momentum contrasts bound into single objects. Their spatial arrangement is predominantly side-by-side across the span, secondarily along the stream and occasionally stacked in height. Their vertical organisation is in the sweep-ejection sense almost everywhere and most active at $y^+=40$--$100$. The importance peaks on the $u=0$ surface, where the fluctuation vanishes, and not in the lobe interiors, where it is largest, so the attribution is not simply tracking fluctuation amplitude (Fig.~\ref{fig3}, right column).

The classical picture of wall turbulence, in which the flow divides into uniform-momentum zones separated by thin inclined layers of concentrated shear, was assembled almost entirely from measurements in streamwise--wall-normal planes~\cite{meinhart_existence_1995, silva_uniform_2016, silva_interfaces_2017}, so only the streamwise and stack arrangements were visible. That picture now emerges as a planar slice of a three-dimensional interface network, and the arrangement it shows, the flat wall-parallel interface between stacked momentum zones, is a small statistical minority. The most common configuration is instead the near-vertical spanwise interface, with its streamwise-vorticity sheet and the steepest jump of the three interfaces. That object is not new to the literature, where it appears piecewise as spanwise-adjacent sweep-ejection pairs in channels, streak-like interfacial layers detected in three dimensions and azimuthal swirls flanking internal shear layers in pipes~\cite{lozano-duran_three-dimensional_2012, dong_coherent_2017, fan_detection_2019, gul_internal_2020, chen_uniform-momentum_2021}. What has been missing is their relative weight, which the streamwise-plane view could not supply.

On the lift-up interpretation reported for the spanwise pairs, the interface is the lateral boundary of the slow lobe and the vorticity sheet is the roll that raises it: one structure seen either as a surface or as a swirl. Because the spanwise-inflectional profile across such a boundary is unstable, the arrangement plausibly closes on itself, which would make it the spanwise-periodic core of the self-sustaining process \cite{hamilton_regeneration_1995, waleffe_self-sustaining_1997, montemuro_self-sustaining_2020}.

Lift-up accounts for the arrangement without establishing what the object is, and the overlap statistics settle that. Every pair arrangement sits about half on Q events and between a third and a half on streaks, with none of the six favouring either family (Supplementary Note~8). A pair therefore holds part of both and is a relabelling of neither. That is the loop the classical comparison closes: the importance structures inherit the kinematic definition of streaks at weak APG, absorb the intense Q events as the lobes of the pairs, ignore the vortex clusters, and add the connecting interface fluid that no kinematic-based threshold retains. Under deceleration, all families converge towards the same compact proportions while steepening.

These observations come from a single wing section at one chord Reynolds number and one angle of attack, so the universality of the $55:38:5$ split and of the two-thirds budgets will require further verification. The friction Reynolds number is also modest, $Re_\tau\approx120$ to $250$, so the inner and outer layers are not well separated: the $y^+=40$--$100$ band where the pairs are most organised sits where the two still overlap, and a larger scale separation would resolve them more cleanly. A further limitation is specific to a decelerating flow. On a wing the pressure gradient rises monotonically along the chord, so the local $\beta$, the accumulated pressure-gradient history and the streamwise position cannot be told apart within one geometry. Boundary layers that reach the same $\beta$ by different routes are known to differ in their mean and fluctuating profiles \cite{bobke_history_2017, vinuesa_revisiting_2017}. Every quantity reported here is normalised by locally measured scales, which absorbs part of that history. Even so, the numeric thresholds quoted along the ramp, and the crossover near $\beta\approx1$ in particular, belong to this $\beta$ distribution rather than to $\beta$ as such.

The results presented here have important implications for flow control: as the APG grows the most informative locations become the momentum interfaces, so sensor placement guided by the census would target interface fluid, not intense events. Given that the most common interface stands across the span, streamwise arrays built for streaks would miss those relevant regions. Because the interfaces keep their viscous-unit geometry while their strength scales, to within a factor of order one, with the local turbulence, an interface-aware wall model would need a viscous length scale and a velocity scale set by that turbulence. Cross-plane particle-image velocimetry can observe the spanwise interface directly, with Fig.~\ref{fig3}d as a template, and the prediction to check is concrete: the spanwise arrangement holds a larger interface volume and a steeper jump than the streamwise one.

The present study also makes a methodological point. An attribution field from a predictive model, segmented and classified with the same statistical care as any physical field, can discover organising structures: it reproduces the known ones quantitatively (shear layers, zone steps, burst fronts) and ranks them together with new ones by predictive relevance while remaining fully auditable. This marks a shift from deep-learning models used as black-box predictors to interpretable frameworks in which the physics is read from the data rather than prescribed.

%%==================================%%
%%==================================%%
\section{Methods}
\label{sec:methodology}

% \begin{figure}[ht!]
%     \centering
%     \includegraphics[width=1\linewidth]{figures/Methodology.png}
%     \caption{Methodology}
%     \label{fig:methodology}
% \end{figure}

%%==================================%%
\subsection{Flow and dataset}
\label{sec:simulation}

The analysis is based on a high-resolution large-eddy simulation (LES) of the turbulent boundary layer developing over the suction side of a NACA~4412 wing section, at chord Reynolds number $Re_c=U_\infty c/\nu=2\times10^5$, based on the freestream velocity $U_\infty$ and the chord length $c$, and at $5^\circ$ angle of attack. Transition is tripped at $x/c=0.1$ on both sides \cite{schlatter_turbulent_2012}. The flow is computed with the spectral-element solver \textsc{Nek5000} \cite{fischer_nek5000_2008}, and the setup was validated against the corresponding fully resolved direct numerical simulation \cite{hosseini_direct_2016, negi_unsteady_2018}. Full numerical details are given by Vinuesa et al.~\cite{vinuesa_turbulent_2018}. This configuration develops an adverse pressure gradient representative of realistic aerodynamic conditions \cite{vinuesa_pressure-gradient_2017, bobke_history_2017, tanarro_effect_2020}. Superscripts $+$ used throughout indicate quantities normalised in viscous units, that is, scaled with the friction velocity $u_\tau=\sqrt{\tau_w/\rho}$ or the viscous length $\ell^*=\nu/u_\tau$, where $\tau_w$ is the wall-shear stress, $\rho$ the fluid density and $\nu$ the kinematic viscosity. Viscous scales are always calculated locally at each streamwise location, unless specified otherwise.
 
The database contains 10{,}000 snapshots, the stored instantaneous fields of the velocity fluctuations $\mathbf{u}=(u,v,w)$ along the wall-parallel, wall-normal and spanwise directions of the suction side. They are sampled every $\Delta t^+=5$ viscous time units, referred to the point of maximum curvature at $x/c=0.4$. The database covers roughly $340$ eddy turnover times $T_e = L_e / u_\tau$ at the same $x/c$ location, the time a swirling structure of size $L_e$ takes to travel its own length or to complete one rotation~\cite{pope_turbulent_2000}. Before the machine-learning analysis, the fields are spectrally interpolated onto a structured grid of $575\times271\times285$ points aligned with the local wall-tangential ($t$) and wall-normal ($n$) directions. The grid preserves the original resolution of the high-resolution LES, with $\Delta x_t^+\approx18$, $\Delta y_n^+\approx0.5$--$11$ and $\Delta z^+\approx9$ at $x/c=0.4$ \cite{choi_grid-point_2012}. The construction and preparation of the database are described in Refs.~\cite{molina-casino_identification_2026, molina-casino_uncovering_2026}.
 
The analysis covers the region $0.25<x/c<0.86$ outlined in Fig.~\ref{fig1}b, where the Clauser pressure-gradient parameter grows from $\beta\approx0.1$ to $\beta\approx10$ and the friction Reynolds number from $Re_\tau=u_\tau \delta_{99}/\nu\approx120$ to $250$, with $\delta_{99}$ the boundary-layer thickness. The lower bound leaves the tripping behind. Beyond the upper bound the structures detach from the wall and the boundary-layer description loses its meaning, so earlier studies of this wing section have used the same range \cite{atzori_coherent_2020, atzori_control_2022}. Because $\beta$, $Re_\tau$ and $\delta_{99}$ all vary along the chord, stations are compared through locally normalised quantities rather than absolute values. Spanwise energy spectra confirm that the sampling interval resolves the motions associated with both the inner and the outer spectral peak over the full range of $\beta$ (Supplementary Table~1 and Fig.~1).

% \begin{figure}[ht!]
%     \centering
%     \includegraphics[width=\linewidth]{figures/Database.png}
%     \caption{\textbf{a}, Simulation setup, including the transition strip, spectral-element mesh, and velocity field in the interpolation domain.
%     \textbf{b}, Streamwise evolution of the Clauser pressure-gradient parameter.
%     \textbf{c}, Spanwise energy spectra at a near-equilibrium station ($x/c=0.4$) and a strong-APG station ($x/c=0.85$). Quantities are expressed in local viscous units.}
%     \label{fig:database}
% \end{figure}

%%==================================%%
\subsection{Predictive model and SHAP attribution}
\label{sec:surrogate_shap}

The analysis follows the explainable deep-learning framework of Cremades et al.~\cite{cremades_classically_2025}, sketched in Fig.~\ref{fig1}a, and extends it to a boundary layer over a wing section \cite{molina-casino_identification_2026, molina-casino_uncovering_2026}. Its premise is that a model able to generate the future state of a turbulent flow bases its predictions on the most influential regions for turbulence transport; feature attribution then recovers those regions~\cite{cremades_additive-feature-attribution_2025}.

A three-dimensional U-Net encoder--decoder network \cite{ronneberger_u-net_2015} based on convolutional layers \cite{lecun_handwritten_1989} is trained to predict $\mathbf{u}(t^++\Delta t^+)$ from $\mathbf{u}(t^+)$, one stored snapshot ahead. Training uses 7{,}500 snapshots with an 80/20 training--validation split and a mean-squared-error (MSE) loss, with the remaining 2{,}500 snapshots reserved for testing. Training is stopped when the relative error inside the boundary layer is about $1\%$ for each velocity component. Accuracy is measured only on the 2{,}500 fields the network never saw. Once trained, the model serves as a differentiable surrogate of the flow evolution, and the additive attribution is computed on it to identify the regions most influential for that evolution. The attribution is run on 6{,}000 snapshots drawn from the whole database.
%Prediction accuracy is not an end in itself here: the model serves as a differentiable surrogate of the flow evolution on which the attribution is computed. The attribution reported below is run on 6{,}000 snapshots drawn from the whole database, because it exploits the trained model rather than testing it. Once the model is known to generalise, the Shapley values of a field are a property of the model evaluated at that field, not a performance estimate.

Input and output data are interpolated from the original flow field onto a Cartesian frame that follows the curvature of the suction side. The grid points are arranged as a $575\times271\times285$ tensor in the streamwise, wall-normal and spanwise directions, with the three velocity fluctuations as channels; each component is rescaled to $[0,1]$ by its extrema over the database before training and mapped back afterwards. The repeated unit is a block made of a $3\times3\times3$ convolution at unit stride with ``same'' padding, which leaves the grid size unchanged, followed by batch normalisation and a rectified-linear activation. The encoder stacks four levels of two blocks each, carrying 24, 48, 96 and 192 filters, separated by $2\times2\times2$ average pooling that halves every dimension. The decoder mirrors that path: each of its levels begins with a $3\times3\times3$ transposed convolution of stride 2 that doubles the resolution, concatenates the result with the encoder level of matching size through a skip connection, and applies one further block. The skip connections return the fine detail that pooling discards, which is what lets the network output a field at the full resolution of its input. A final $3\times3\times3$ convolution to three channels with a sigmoid activation returns the normalised prediction. Altogether the network has fifteen convolutional layers and $3.3\times10^6$ trainable parameters, and its loss is minimised with RMSprop at a learning rate of $10^{-4}$ and a momentum of $0.9$, in single precision and with one field per batch due to memory constraints. 
 
The contribution of each input location to the prediction is quantified with SHapley Additive exPlanations (SHAP) \cite{shapley_value_1953, lundberg_unified_2017}. The prediction is treated as a cooperative game in which every grid point is a player, and the importance vector at point $i$, $\mathbf{\Phi}_i=(\phi_{i,u},\phi_{i,v},\phi_{i,w})$, is its average marginal contribution over all subsets of inputs. An exact evaluation would have a computational cost of $\mathcal{O}(2^{10^7})$, which would be prohibitive, so the Shapley values are approximated with gradient-SHAP \cite{erion_improving_2021}:
% \begin{strip}
\begin{equation}
    \label{eq:SHAP}
    \phi_i(x_{in})=
    \mathbb{E}_{x_{ref},\alpha\sim U(0,1)}
    \left[
    (x_{in,i}-x_{ref,i}) \frac{\partial F(x_{ref}+\alpha(x_{in}-x_{ref}))} {\partial x_{in,i}}
    \right],
\end{equation}
% \end{strip}
where $F$ is the model evaluating the MSE of the prediction and $\alpha$ interpolates between the reference field $x_{ref}$ and the input field $x_{in}$. The reference is the zero-fluctuation field, which represents the mean boundary layer and provides a physically meaningful baseline. The magnitude $|\mathbf{\Phi}|=\sqrt{\phi_u^2+\phi_v^2+\phi_w^2}$ is used throughout as the scalar measure of local dynamical importance.

%%==================================%%
\subsection{Structure detection}
\label{sec:percolation}

All structures, importance-based and classical, are detected from the same instantaneous fields and with the same procedure, so that the comparison of Fig.~\ref{fig2} is not biased by the detection itself.
 
The domain is first restricted to the turbulent region with a turbulent/non-turbulent interface (TNTI) mask \cite{atzori_coherent_2020}, which retains the points where the instantaneous turbulent kinetic energy, $\tilde{k}$, exceeds $45\%$ of its value at the boundary-layer edge, $\tilde{k}>0.45\,k_e$. The boundary-layer edge is obtained from the diagnostic-plot method \cite{alfredsson_new_2011, vinuesa_determining_2016}. Without the TNTI mask, the vanishing fluctuations of the irrotational free stream produce spurious detections above the boundary-layer edge \cite{molina-casino_uncovering_2026}.
 
Within the masked region, a grid point belongs to an importance structure when
\begin{equation}
    \label{eq:shap_struc}
    |\mathbf{\Phi}(\mathbf{x},t)| >
    H_\Phi \sqrt{\overline{\phi_u^2}(x,y) + \overline{\phi_v^2}(x,y) + \overline{\phi_w^2}(x,y)},
\end{equation}
where the overbar denotes averaging in time and in the spanwise direction, and $H_\Phi$ is the percolation threshold on the importance field. Connected components above the threshold define the structures. The threshold is not chosen by hand but obtained from a percolation analysis. As $H_\Phi$ increases, the field passes from a single connected network of voxels to a set of isolated objects, and $H_\Phi^\star$ is taken where the number of detected objects is largest. For the present database this yields $H_\Phi^\star=1.71$. Components whose volume falls below $30^3$ viscous units are discarded at every stage of the analysis, in every family alike, because objects that small have no shape the grid can resolve; what remains is $4.3\times10^6$ importance structures over the 6{,}000 snapshots used for post-processing. Fig.~\ref{fig1}b shows those detected in one instantaneous field.

%Because the reference field of Eq.~(\ref{eq:SHAP}) is the zero-fluctuation state, the attribution carries the local fluctuation as a multiplicative factor, and the threshold of Eq.~(\ref{eq:shap_struc}) is referred to the local root-mean-square of the importance itself. Ratios of an interface amplitude to the local turbulence intensity are therefore partly conditioned by this construction, and are reported here as measurements rather than as scalings.
 
The same detection is applied to the three classical families, each with its own percolation-derived threshold \cite{lozano-duran_three-dimensional_2012}. Intense Reynolds-stress events, or Q events, satisfy $|uv|>H_{uv}\,u'v'$ with $H_{uv}=1.62$, where $u'$ and $v'$ are the local root-mean-square values of the streamwise and wall-normal fluctuations and $H_{uv}$ is the hyperbolic-hole size of quadrant analysis, distinct from the percolation threshold $H_\Phi$ above. They are classified by their quadrant in the $uv$ plane: Q1 (outward: $u>0$, $v>0$), Q2 (ejections: $u<0$, $v>0$), Q3 (inward: $u<0$, $v<0$) and Q4 (sweeps: $u>0$, $v<0$). Streamwise streaks \cite{kline_structure_1967} satisfy $\sqrt{u^2+w^2}>3.81\,u_\tau(x)$ and are classified by the sign of $u$ into high- and low-speed. Vortex clusters are detected with the $\Delta$-criterion of Chong et al.~\cite{chong_general_1990}, $\Delta>0.114\,\Delta'$, where $\Delta$ is the discriminant of the velocity-gradient tensor.

The four criteria are not normalised alike, and the difference matters on a ramp along which the friction velocity falls steeply. The importance, Q-event and vortex thresholds are each referred to a local root-mean-square that varies with chord position and with wall distance, whereas the classical streak criterion is referred to $u_\tau(x)$, a single value per station with no wall-normal dependence, and one that becomes very small as separation approaches~\cite{vinuesa_pressure-gradient_2017}. The streak definition is kept in its classical form, so the comparison is with the streak as the literature defines it and not with a rescaled variant. Streak quantities measured at the strongest-APG stations are therefore conditioned on that choice in a way the other three families are not. %Expressed in local intensity units the streak threshold is therefore $3.81\,u_\tau/u'$, which drifts along the chord as $u'/u_\tau$ grows, while the other three hold their position relative to the local statistics by construction. The streak definition is kept in its classical form, so the comparison is with the streak as the literature defines it and not with a rescaled variant. Streak quantities measured at the strongest-APG stations are therefore conditioned on that choice in a way the other three families' are not.

Every structure also carries a geometry, measured on the same grid and by the same rules for all four families and with no reference to the taxonomy. The bounding dimensions $\Delta x$, $\Delta y$ and $\Delta z$ are the extents of the smallest grid-aligned box that encloses the connected set of points. Dividing a dimension by the local viscous length $\ell^*$ at the structure's centroid puts it in wall units, and the ratio of two dimensions is an aspect ratio that needs no scaling at all. The main text uses the streamwise-to-wall-normal elongation $\Delta x/\Delta y$ (Fig.~\ref{fig2}c, top); the other two ratios, and the joint size distributions behind them, are given in Supplementary Note~7.

How a structure fills that box is measured by the inclination $\theta_{xy}$ (Fig.~\ref{fig2}c, bottom). Each structure is first given its volume-weighted centroid. The straight line joining that centroid to a point inside the structure makes an angle with the wall-parallel direction in the streamwise--wall-normal plane, folded onto $(-90^\circ,90^\circ]$, and $\theta_{xy}$ is the volume-weighted mean of that angle over the structure. A positive $\theta_{xy}$ therefore means that the structure leans away from the wall as it goes downstream.

Agreement between the importance structures and each classical family is measured on the voxel sets. Writing $S$ for the voxel set of the importance structures, the composition $(X\cap S)/S$ is the fraction of the importance volume occupied by family $X$, and the capture $(X\cap S)/X$ the fraction of family $X$ recovered by the importance structures. %Velocity-space comparisons use histograms in viscous and in local units at each station (Supplementary Figs.~15 and 16), and the overlap profiles are given at every reference station in Supplementary Fig.~18. The overlap resolved by class and by pair arrangement is given in Supplementary Figs.~19 and 20.

%%==================================%%
\subsection{Post-processing of the importance structures}
\label{sec:postprocessing}
 
Once detected, each importance structure is classified and the resulting populations are averaged into conditional averages. The taxonomy follows the classification tree of Fig.~\ref{fig1}, whose classes name the dominant motion a structure contains and not a new type of structure. Each structure is tested against four questions in order and assigned to the first branch it satisfies, every test requiring the property to hold over more than $80\%$ of the structure volume. (i) Intensity: if the points satisfy the Q-event criterion given above, the structure is assigned to the pure Q class of its dominant quadrant, pure Q2 (ejection), pure Q4 (sweep), pure Q1 (outward) or pure Q3 (inward). (ii) Streamwise sign: otherwise, if the points share the sign of $u$, the structure is pure HS (high-speed, $u>0$) or pure LS (low-speed, $u<0$). (iii) Wall-normal sign: otherwise, if the points share the sign of $v$, the structure is an updraft ($v>0$) or a downdraft ($v<0$). (iv) All remaining structures mix both signs of $u$ and both signs of $v$ and are labelled HS/LS pairs. Classes (i) to (iii) hold one dominant motion each and are the single-lobed classes; the pairs of (iv) are the double-lobed class. The ordering matters: a structure that would satisfy more than one test is always assigned to the earliest, so the pure Q classes are the strictest and the pair class is defined by exclusion.

Each class is then given five budgets, each the share of a global total held by that class. The first is the number of structures, while the other four sum the volume, the importance $\sum|\mathbf{\Phi}|$, the Reynolds shear stress $\sum|uv|$ and the turbulent kinetic energy $\sum k$ over the voxels of the class, each normalised by the same sum over all classes. Density budgets divide the previous magnitudes by the volume of the class and are renormalised to sum to unity, so that they rank classes by content per unit volume rather than by total content. Every density share quoted in this paper uses the seven classes of the classification tree, pooling Q1 with Q3 and updrafts with downdrafts and keeping the pair class whole. The partition-free form of the same quantity is the ratio of the class density to the population-mean density, which measures how much richer than average the fluid of a class is; the Supplementary tables report both. The census of Fig.~\ref{fig1}c reports the importance budget and its density, resolved along the pressure-gradient ramp.

Each HS/LS pair is then typed by where its two lobes sit relative to one another, which is what the arrangement schematic of Fig.~\ref{fig3}a reports. The measurement is the offset from the low-speed lobe centroid to the high-speed lobe centroid, expressed in local wall units, and its direction is tested in three steps. First, the offset must lie between 5 and 400 wall units; pairs outside that window are left untyped. Second, an offset that falls within a $30^\circ$ half-angle cone about the wall-normal direction makes the pair a vertical stack, one lobe sitting above the other. Third, the offsets that survive that test are wall-parallel, and a $30^\circ$ cone about the streamwise direction splits them in turn: inside the cone the pair is streamwise, one lobe ahead of the other along the flow; outside it the pair is spanwise, the two lobes side by side across the span. The three tests therefore yield three arrangement axes, and because the offset also carries a sign, each axis splits further into two orientations, hence six arrangements: fast lobe downstream or upstream, fast lobe to one side of the span or the other, and fast lobe above or below. Fig.~\ref{fig3}b--e shows four of the six.

The vertical organisation of a pair is scored next; this is the quantity plotted in Fig.~\ref{fig4}b,c. Each lobe, the low-speed ($u<0$) and the high-speed ($u>0$) part of the pair, is reduced to the volume imbalance between its rising ($v>0$) and its sinking ($v<0$) fluid. Writing $V_{v>0}$ and $V_{v<0}$ for those two volumes, the low- and high-speed lobe coherences are
\begin{equation}
    c_- = \left. \frac{V_{v>0} - V_{v<0}}{V_{v>0} + V_{v<0}} \right|_{u<0}, 
    \;
    c_+ = \left. \frac{V_{v>0} - V_{v<0}}{V_{v>0} + V_{v<0}} \right|_{u>0},
\end{equation}
each in $[-1,1]$ and positive when its lobe predominantly rises. sweep-ejection transport asks opposite things of the two lobes: the slow lobe must rise, which is an ejection, and the fast lobe must sink, which is a sweep. A single score is therefore built by taking the lobe with the more one-sided vertical motion, that is, the larger $|c|$, and negating the high-speed coherence so that sweep-ejection motion counts as positive on either side. This is the pair's sweep-ejection alignment,
\begin{equation}
    A = \begin{cases} \;c_- & \text{if } |c_-| \ge |c_+|,\\[2pt] -c_+ & \text{otherwise}.\end{cases}
\label{eq:alignment}
\end{equation}
Thus $A=+1$ is perfect sweep-ejection organisation (Q2/Q4) and $A=-1$ its outward-inward reverse (Q1/Q3), while the magnitude $|A| = \max(|c_-|,|c_+|)$ measures how one-sided the motion in that lobe is. A pair is counted as organised when $|A|$ passes the value implied by the taxonomy's own $80\%$ volume share, and the two are connected by a short calculation: a lobe in which a fraction $f$ of the volume rises has a coherence $c = f - (1-f) = 2f-1$, so the $80\%$ share that makes a structure pure in every other class corresponds to $A = 2\times0.80 - 1 = 0.6$ (cut used in Fig.~\ref{fig4}b,c). Moving that cut moves the level of the resulting fraction, so the main text reads these curves for their trends alone; the median alignment $\tilde{A}$ is reported alongside because it involves no threshold and follows the same trends.

Each defined population is then averaged into a composite (Fig.~\ref{fig3}b--e). The velocity and importance fields are sampled on a fixed grid of wall-unit offsets ($\pm200$ streamwise, $\pm100$ wall-normal, $\pm150$ spanwise) centred on the pair's midpoint. The two spanwise orientations are kept as separate populations rather than mirrored onto each other. Several quantities are measured on each composite: the interface line (fitted $u=0$ line) and its angle, the 10--90\% velocity jump and thickness along the interface normal, the amplitude of the vorticity sheet, and the aspect ratio of the region enclosed by the 90th percentile of the importance. The composites of Fig.~\ref{fig5} repeat the whole procedure within four bands of $\beta$ ($\beta<1$, $1$--$5$, $5$--$10$, $\ge10$), while the curves of Fig.~\ref{fig5}c are measured in narrower $\beta$ bins along the chord.

Because the census of Fig.~\ref{fig1}c and the pair typing of Fig.~\ref{fig3}a rest on the $80\%$ classification threshold, which could in principle be set otherwise, its sensitivity is checked directly. The classes are not equally separated in velocity space: a structure of slow fluid that also rises coherently lies close to the boundary between pure LS and pure Q2 (ejection), so lowering the threshold transfers volume from the former to the latter. Sweeping the threshold over $60\%$--$95\%$ therefore shifts the levels, mainly by exchange between adjacent classes. It changes neither the shape nor the sign of any trend, and HS/LS pairs remain the dominant class at every value (Supplementary Note~2). The percentages quoted for the census should therefore be read as values obtained at 80\%, not as threshold-independent constants.

% %%==================================%%
% %%==================================%%
% \section{Conclusions}\label{sec:conclusions}

% Sample body text. Sample body text. Sample body text. Sample body text. Sample body text. Sample body text. Sample body text. Sample body text.

%%==================================%%
%%==================================%%
\backmatter
\bookmarksetup{startatroot}

\bmhead{Supplementary information}

This article is accompanied by Supplementary Information containing Supplementary Notes~1--8, Supplementary Figures~1--20 and Supplementary Tables~1--5.

\bmhead{Acknowledgements}

The authors gratefully acknowledge the computer resources at MareNostrum provided by the Barcelona Supercomputing Center (IM-2025-1-0002). 
This project was provided with HPC computing and storage resources by GENCI at IDRIS thanks to the grant 2024\_AD012A15788 on the supercomputer Jean Zay's CSL partition; at TGCC thanks to the grants 2024\_AD012A16003 and 2025\_AD012A16168R1 on the supercomputer Joliot-Curie's Irene SKL/Rome partition; and at IDRIS thanks to the grant 2025\_AD011016855 on the supercomputer Jean Zay's H100 partitions. 
Part of the calculations was enabled by resources provided by the National Academic Infrastructure for Supercomputing in Sweden (NAISS) on Alvis at C3SE. 
The authors used Claude (claude.ai) to assist with grammar checking, spelling and readability. All generated text was thoroughly reviewed, edited and validated by the authors, who take full responsibility for the content.

\section{Declarations}

\bmhead{Funding}

S. M.-C. acknowledges funding from the Scientific Directorate of ONERA and from ERC grant no. 2021-CoG-101043998, DEEPCONTROL.
R. V. acknowledges the financial support from ERC grant no. 2021-CoG-101043998, DEEPCONTROL. Views and opinions expressed are however those of the authors only and do not necessarily reflect those of the European Union or the European Research Council. Neither the European Union nor the granting authority can be held responsible for them. 
The data has been obtained with support from MCIN/AEI/10.13039/501100011033 and by ERDF, ``A way of making Europe'', under project PID2024-162480OB-I00 and CIPROM/2024/35 by GVA (S. H.). 

\bmhead{Author contributions}

S. M.-C.: Conceptualization, Writing – original draft, Visualization, Validation, Software, Methodology, Investigation. 
A. C.: Writing – review \& editing, Software, Methodology, Investigation, Data curation. 
S. H.: Writing – review \& editing, Methodology, Investigation, Data curation, Resources, Funding acquisition. 
J. I. C.: Writing – review \& editing, Investigation, Supervision, Funding acquisition. 
F. C.: Writing – review \& editing, Investigation, Supervision. 
R. V.: Initial idea, Writing – review \& editing, Project administration, Methodology, Funding acquisition, Conceptualization.

\bmhead{Competing interests}

The authors declare no competing interests.

\bmhead{Data availability}

The minimum representative downsampled data used in this study will be made available open access as soon as the article is published. For the complete database, please contact the authors.

\bmhead{Code availability}

The codes used to produce this study will be made available for open access as soon as the article is published.

% %%===================================================%%
% %% For presentation purpose, we have included        %%
% %% \bigskip command. Please ignore this.             %%
% %%===================================================%%
% \bigskip

% Springer journals and proceedings: \url{https://www.springer.com/gp/editorial-policies}

% Nature Portfolio journals: \url{https://www.nature.com/nature-research/editorial-policies}

% \textit{Scientific Reports}: \url{https://www.nature.com/srep/journal-policies/editorial-policies}

% BMC journals: \url{https://www.biomedcentral.com/getpublished/editorial-policies}
% \end{flushleft}


%%===========================================================================================%%

\bibliography{references}% common bib file
% \input sn-article.bbl

\clearpage
\setcounter{page}{1}
% \bookmarksetup{startatroot}
\phantomsection
\pdfbookmark[1]{Supplementary Information}{sec:supplementary}
\renewcommand{\thepage}{S. \arabic{page}}

\begin{appendices}

\input{supplementary}

% \section{Section title of first appendix}\label{secA1}

% An appendix contains supplementary information that is not an essential part of the text itself but which may be helpful in providing a more comprehensive understanding of the research problem or it is information that is too cumbersome to be included in the body of the paper.

% %%=============================================%%
% %% For submissions to Nature Portfolio Journals %%
% %% please use the heading ``Extended Data''.   %%
% %%=============================================%%

\end{appendices}

\end{document}

%% file: supplementary.tex
\graphicspath{{figures/supplementary/}}
\renewcommand{\figurename}{Supplementary Fig.}
\renewcommand{\tablename}{Supplementary table}

% \subsection{Supplementary Note 1: the database, the prediction horizon and the surrogate model}
\subsection{Supplementary Note 1: the prediction horizon}
% =====================================================================================

% The attribution field inherits every property of the model that produced it, so the pipeline is only as trustworthy as the two choices behind that model: how far ahead the network is asked to predict, and how the network is built and trained. Both are fixed once and used everywhere, and both are documented here.

% \paragraph{The prediction horizon.} 
The model predicts the velocity field one stored snapshot ahead. In the units of the simulation that interval is $\Delta t = 6\times10^{-3}\,c/U_\infty$, and it is the same everywhere; expressed in local viscous units it is not, because the friction velocity collapses along the chord. Supplementary Table~\ref{stab:spectra} gives $\Delta t^+ = \Delta t\, u_\tau^2/\nu$ station by station: it falls from $5.7$ at $x/c=0.30$ to $1.0$ at $x/c=0.85$. A horizon that shrinks in wall units as the pressure gradient grows works in our favour, but $\Delta t^+$ on its own is not the number that matters. What matters is whether the horizon is short compared with the turnover time of the motions the model has to track.

\begin{figure*}[ht!]
    \centering
    \includegraphics[width=100mm]{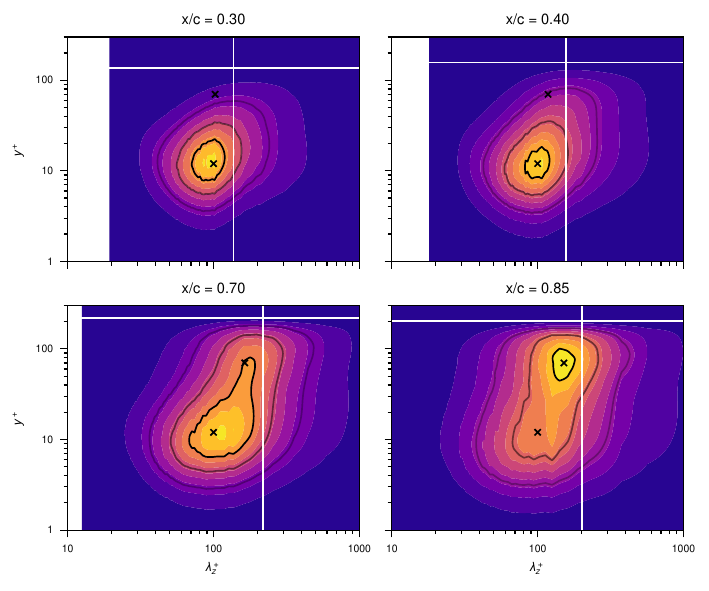}
    \caption{\textbf{Spanwise energy spectra of the streamwise velocity fluctuation at four chord stations.} Premultiplied spanwise spectrum of the streamwise velocity fluctuation, $k_z E_{uu}/u_\tau^2$, against spanwise wavelength $\lambda_z^+$ (horizontal) and wall distance $y^+$ (vertical), at four reference stations (left to right: $x/c=0.30$, $0.40$, $0.70$, $0.85$). Colour, spectral density on a linear scale, dark to bright; black lines, iso-contours. White lines mark $\delta_{99}^+$ on both axes. The crosses mark the reference inner peak ($y^+=12$, $\lambda_z^+=100$) and the outer peak ($y^+=70$, $\lambda_z^+=0.75\,\delta_{99}^+$) whose turnover times are listed in Supplementary Table~\ref{stab:spectra}.}
    \label{sfig:spectra}
\end{figure*}

Supplementary Fig.~\ref{sfig:spectra} answers that with the spanwise energy spectra. The premultiplied spectrum of the streamwise fluctuation shows the two energetic scales of a wall-bounded layer, the inner peak near $y^+\approx12$ at a spanwise wavelength $\lambda_z^+\approx100$, and the outer peak near $y^+\approx70$ at $\lambda_z^+\approx0.75\,\delta_{99}^+$. The two peaks lie close together because the Reynolds number of this flow is relatively low. Dividing each wavelength by the local convection velocity at its own height gives a turnover time for that motion, listed in Supplementary Table~\ref{stab:spectra}. The inner peak turns over in $\Delta t^+\approx10$--$11$ and the outer peak in $\Delta t^+\approx6$--$10$ over the whole chord, against a sampling interval of $5.7$ falling to $1.0$. The horizon therefore matches the outer turnover time at the first station, where it is already half the inner one, and falls to a tenth of both by $x/c=0.85$. The margin widens monotonically downstream, that is, towards the stations where the structural reorganisation happens. The model is asked to advance the flow by a fraction of an eddy lifetime rather than to forecast across it.

The spectra also record the physical change the census of the main text tracks. Between $x/c=0.30$ and $x/c=0.85$ the energy migrates from the inner peak to the outer one, and by $x/c=0.85$ the outer peak dominates. This is the known energisation of the outer layer under an adverse pressure gradient, and it is the mean-flow counterpart of the structural takeover reported in main-text Fig.~1c.

\begin{table*}[ht!]
    \centering
    \caption{\textbf{The prediction horizon against the turnover time of the energetic motions.} $\Delta t^+$ is the fixed sampling interval $\Delta t = 6.11\times10^{-3}\,c/U_\infty$ expressed in the viscous units of each station. The turnover times are $\lambda_z^+/U_c^+$, the spanwise wavelength of each spectral peak divided by the mean velocity at the height of that peak, with $U_c^+$ read from the mean profile. The inner-peak wavelength is fixed at the canonical $\lambda_z^+=100$; the outer-peak wavelength scales with the local boundary-layer thickness and is listed.}
    \label{stab:spectra}
    \vspace{4pt}
    \begin{tabular}{lrrrrrrr}
        \toprule
         & & & \multicolumn{2}{c}{inner peak} & \multicolumn{3}{c}{outer peak} \\
        \cmidrule(lr){4-5}\cmidrule(lr){6-8}
        $x/c$ & $\beta$ & $\Delta t^+$ & $U_c^+$ & $\Delta t_t^+$ & $\lambda_z^+$ & $U_c^+$ & $\Delta t_t^+$ \\
        \midrule
        0.30 & 0.3  & 5.7 &  9.8 & 10.2 & 103 & 18.0 & 5.7 \\
        0.40 & 0.7  & 5.0 &  9.4 & 10.7 & 118 & 16.8 & 7.0 \\
        0.70 & 2.7  & 2.4 &  9.4 & 10.6 & 164 & 16.5 & 9.9 \\
        0.85 & 9.8  & 1.0 & 10.3 &  9.7 & 151 & 20.6 & 7.4 \\
        \bottomrule
    \end{tabular}
\end{table*}

\clearpage
% =====================================================================================
\subsection{Supplementary Note 2: the classification, its validation and its robustness}
% =====================================================================================

The classification tree of main-text Fig.~1 names every importance structure from its own velocity content, using a single free parameter: the $80\%$ volume share that each test has to satisfy.

\begin{figure*}[ht!]
    \centering
    \includegraphics[width=\linewidth]{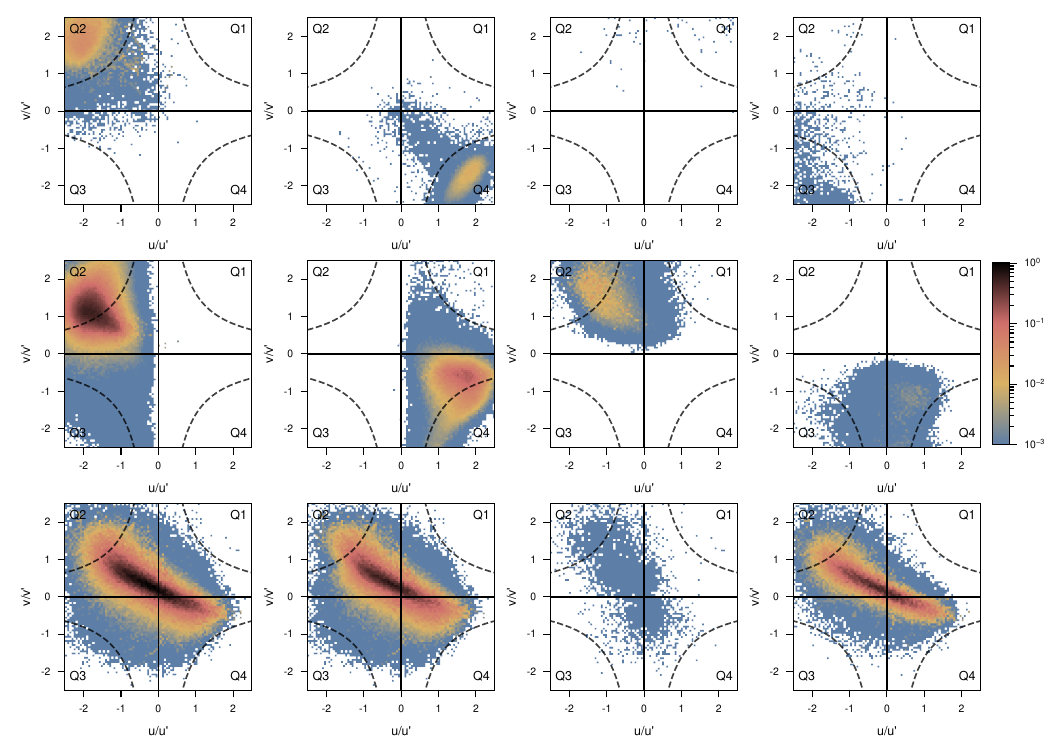}
    \caption{\textbf{The classes of the taxonomy in the plane of the velocity fluctuations.} Volume-weighted density of the structure fluctuations $(u/u',v/v')$, all points of each structure considered, one panel per class, on a common logarithmic colour scale. Dashed curves, the hyperbolic hole $|uv|=1.62\,u'v'$ that delimits the intense quadrant events; quadrant labels in the corners. Top row, the four pure Q classes: pure Q2 (ejection), pure Q4 (sweep), pure Q1 (outward) and pure Q3 (inward). Middle row, the single-signed classes: pure LS, pure HS, updrafts and downdrafts. Bottom row, the pairs, first pooled and then split into the sweep-ejection, outward-inward and vertically unorganised groups. The pure Q1, pure Q3 and outward-inward pair panels are sparse because those classes are rare, not because their fluid is spread out.}
    \label{sfig:quadrant}
\end{figure*}

\paragraph{The census is coherent in the quadrant plane.} A classification built from volume fractions could in principle produce classes that are bookkeeping artefacts rather than physical populations. The test is to map each class onto a space it was not defined in. Supplementary Fig.~\ref{sfig:quadrant} places every class in the quadrant plane of the structure fluctuations, $(u/u',v/v')$, together with the hyperbolic hole $|uv| = 1.62\,u'v'$ that separates intense quadrant events from moderate-amplitude fluid. Each class falls where its definition requires, and several of them fall in a way the definition never enforced. The pure Q2 and pure Q4 structures sit deep inside the Q2 and Q4 branches, outside the hole, which their intensity test does demand, and their outward-inward (counter-gradient) counterparts, pure Q1 and pure Q3, are confined to their own branches in the same way, though they are far rarer. The pure LS and pure HS classes are confined to their own sign of $u$, which their test also demands. Nothing in the tree constrains their vertical velocity, yet they concentrate on the sweep-ejection (down-gradient) side: pure LS fluid leans into the ejection quadrant and pure HS fluid into the sweep quadrant. That lean is the correlation between speed and wall-normal motion that the lift-up effect produces, recovered here as a property of the classes rather than imposed on them. The updrafts and downdrafts, whose test is on the sign of $v$, lie entirely on one side of the $v$ axis and straddle both speeds. The pairs run along the sweep-ejection diagonal and pass straight through the hole.

That last row is the one the main text uses twice. It shows why no intensity criterion can find the HS/LS pairs: the fluid that binds the two lobes of a pair lives in the region that quadrant analysis discards by construction. Splitting the pairs by their sweep-ejection alignment sharpens the picture. The sweep-ejection group reaches into both the ejection and the sweep branch while keeping most of its mass inside the hole. The vertically unorganised group lies along the same diagonal but much closer to the $u$ axis, its mean vertical velocity small. The outward-inward group is a sparse cloud around the origin, with no preferred direction and a density three orders of magnitude below the other two. The three are not three kinds of object; they are three slices of one continuous distribution.

\begin{figure*}[ht!]
    \centering
    \includegraphics[width=120mm]{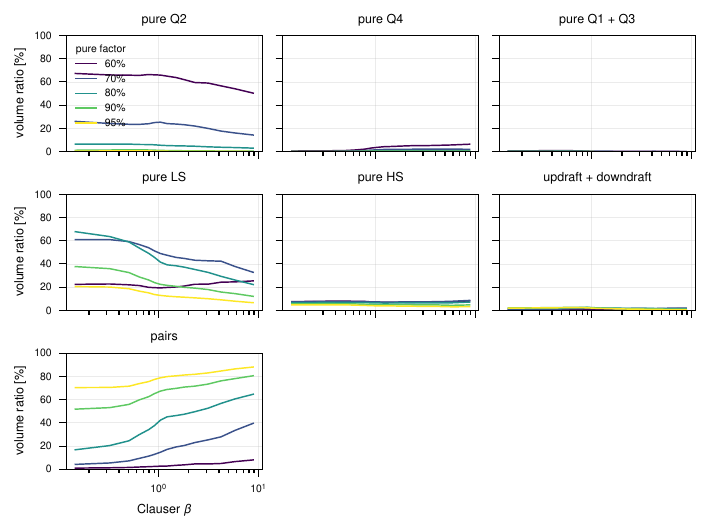}
    \caption{\textbf{The census repeated at five values of the classification threshold.} Volume share of each class against Clauser $\beta$, one panel per class, for five values of the single classification threshold (60\%, 70\%, 80\%, 90\%, 95\%; legend in the first panel).}
    \label{sfig:threshold}
\end{figure*}

\paragraph{The conclusions survive a different threshold.} The 80\% share is a methodological choice, and the classes it separates are not equally distinct. A structure of slow fluid that also rises coherently sits close to the boundary between pure LS and pure Q2 (ejection), so lowering the share moves volume from the first class into the second. Supplementary Fig.~\ref{sfig:threshold} repeats the entire classification at shares of 60\%, 70\%, 80\%, 90\% and 95\% and follows each class along the pressure-gradient ramp.

The levels move most between neighbouring classes: at a share of 60\% more than half of the volume is detected as pure Q2 (ejection), at 95\% almost none of it is. What does not move is the shape or the sign of any trend. Pure LS structures fall monotonically with $\beta$ at every threshold, the pairs rise monotonically at every threshold, the outward-inward classes stay negligible at every threshold, and the crossover between the two dominant classes stays near $\beta\approx1$. The pairs remain the largest class by volume at $\beta\approx10$ for every share of 70\% and above, and at a share of 60\% they are displaced only because the pure Q2 class has absorbed most of the population. The percentages quoted in the main text should therefore be read as values obtained at 80\%, and the trends as threshold-independent.

\clearpage
% =====================================================================================
\subsection{Supplementary Note 3: the census in full}
% =====================================================================================

Main-text Fig.~1c compresses the census into one budget and its density, both as shares. The full census is reported here: five budgets, each pooled over the population and each resolved along the chord, in absolute value as well as in share. The absolute curves answer a question the shares cannot, namely whether a falling share means a shrinking class or a growing flow.

\paragraph{The pooled census.} Supplementary Table~\ref{stab:census_absolute} gives the share of the structure count, the volume, the importance $\sum|\mathbf{\Phi}|$, the Reynolds shear stress $\sum|uv|$ and the turbulent kinetic energy $\sum k$ held by each class. The pattern is the one the main text states and is worth seeing in full. Structures made almost entirely of pure Q2 (ejection) or pure Q4 (sweep) fluid are numerous, $14\%$ of the objects, and small: together they hold $4.5\%$ of the volume, $2.1\%$ of the importance and $4.0\%$ of the stress. Pure LS structures are the most numerous class, $45\%$ of the objects, and hold about a third of every budget. The pairs run the other way, and are the one class for which the count-weighted and the volume-weighted views disagree. They are one object in six, $16.7\%$ of the count, yet they hold $52.6\%$ of the volume, $49.8\%$ of the importance, $47.2\%$ of the stress and $47.4\%$ of the energy: half of every budget, carried by a sixth of the population. Every single-lobed class holds a smaller share of the volume than of the count; only the pairs hold more, which makes an average pair about three times the volume of an average structure (last column of Supplementary Table~\ref{stab:census_absolute}). That size gap is what lets a pair hold intense fluid without being built from it: the intense fluid sits in the two lobes, and the volume that carries the class to the top of every budget is the moderate fluid that binds them. The outward-inward classes never reach one percent of any budget, so the reorganisation reported in the main text happens entirely inside the sweep-ejection motions.

Supplementary Table~\ref{stab:census_density} divides each budget by the volume of its class, which asks which fluid is richest rather than which class holds most. The ranking inverts, and the last three columns give the size of the inversion directly as an enrichment factor over the population mean.

One property of a density share has to be stated before the numbers are read, because it governs how they may be compared. The share of Supplementary Table~\ref{stab:census_absolute} is a fraction of a fixed global total, so splitting a class into sub-classes divides that class' entry and leaves every other entry untouched. A density share is not of that kind: it normalises the density of a class by the sum of the class densities, and that sum grows whenever the population is cut more finely, which lowers every other entry of the column. A density share therefore ranks the classes reliably, but its level belongs to the class list as much as to the flow, and levels obtained from different class lists cannot be set against each other. Supplementary Table~\ref{stab:census_density} accordingly uses the seven classes of the main text's classification tree, the same partition as main-text Fig.~1c. The pooled values of the table and the station-resolved values of Fig.~1c are then two views of one quantity, and the first lies inside the range spanned by the second. The enrichment factors of the last three columns are free of the ambiguity altogether, because their denominator is the population-mean density, which no regrouping changes, and they are the form to quote and to compare.

Read that way the table says the following. Pure HS fluid carries $2.1$ times the mean importance density and $2.0$ times the mean energy density, the largest values in the table. Pure Q4 fluid carries $2.1$ times the mean stress density. Pure Q2 fluid, by contrast, is dilute in every budget, at $0.38$ of the mean importance density: ejections are numerous, not concentrated. The pairs sit close to the mean by construction, since they contain both kinds of fluid, at $0.95$ of the mean importance density. Inside the class the sweep-ejection group reaches $1.04$, the vertically unorganised group $0.78$ and the rare outward-inward group $0.47$, so no group departs far from the mean either. Point for point the most informative fluid in this flow is fast, sweep-side fluid, even though slow fluid fills more of the important volume.

\begin{table*}[ht!]
    \centering
    \caption{\textbf{The pooled census.} Share of each budget held by each class, in per cent of the total over all structures, pooled over the whole population and the whole chord. Classes are those of the main text's classification tree. The last column is not a share but a ratio of the first two, the volume share divided by the count share, which gives the mean volume of a structure of that class in units of the population-mean structure volume: a value above one marks a class whose objects are larger than average. It is left blank for pure Q1 and pure Q3, whose count and volume shares are both too small for the ratio to survive their rounding.}
    \label{stab:census_absolute}
    \vspace{4pt}
    \begin{tabular}{lrrrrrr}
        \toprule
        class & count & volume & $\sum|\mathbf{\Phi}|$ & $\sum|uv|$ & $\sum k$ & mean size \\
        \midrule
        pure Q2 (ejection)         & 12.04 &  4.01 &  1.54 &  3.07 &  1.91 & 0.33 \\
        pure Q4 (sweep)            &  1.90 &  0.43 &  0.52 &  0.91 &  0.54 & 0.23 \\
        pure Q1 (outward)          &  0.05 &  0.01 & $<0.01$ & $<0.01$ & $<0.01$ & --- \\
        pure Q3 (inward)           &  0.12 &  0.03 & $<0.01$ & $<0.01$ & $<0.01$ & --- \\
        pure HS                    & 20.50 &  7.35 & 15.79 & 12.02 & 14.99 & 0.36 \\
        pure LS                    & 44.67 & 33.94 & 31.25 & 35.71 & 34.15 & 0.76 \\
        updraft                    &  2.63 &  1.39 &  0.91 &  0.93 &  0.87 & 0.53 \\
        downdraft                  &  1.38 &  0.29 &  0.21 &  0.17 &  0.18 & 0.21 \\
        \textbf{HS/LS pairs}   & \textbf{16.72} & \textbf{52.55} & \textbf{49.78} & \textbf{47.19} & \textbf{47.36} & \textbf{3.14} \\
        \bottomrule
    \end{tabular}
\end{table*}

\begin{table*}[ht!]
    \centering
    \caption{\textbf{The census by density.} Each budget divided by the volume of its class. The first three columns give that density as a share of the summed class densities, in per cent, so they rank the classes; the last three give it as an enrichment factor over the population-mean density, so a value above one marks fluid that is richer than average. The class list is the seven classes of the main text's classification tree, the partition used in main-text Fig.~1c, with the pairs kept whole.}
    \label{stab:census_density}
    \vspace{4pt}
    \begin{tabular}{lrrrrrr}
        \toprule
         & \multicolumn{3}{c}{share of density (\%)} & \multicolumn{3}{c}{enrichment over the mean} \\
        \cmidrule(lr){2-4}\cmidrule(lr){5-7}
        class & $|\mathbf{\Phi}|/V$ & $|uv|/V$ & $k/V$ & $|\mathbf{\Phi}|/V$ & $|uv|/V$ & $k/V$ \\
        \midrule
        pure Q2 (ejection)            &  6.05 & 10.54 &  7.45 & 0.38 & 0.76 & 0.48 \\
        pure Q4 (sweep)               & 19.13 & 29.52 & 19.73 & 1.21 & 2.14 & 1.26 \\
        pure Q1/Q3                    &  1.05 &  1.47 &  1.30 & 0.07 & 0.11 & 0.08 \\
        pure HS                       & 33.86 & 22.55 & 31.93 & 2.15 & 1.64 & 2.04 \\
        pure LS                       & 14.50 & 14.51 & 15.75 & 0.92 & 1.05 & 1.01 \\
        updraft/downdraft             & 10.49 &  9.03 &  9.74 & 0.67 & 0.65 & 0.62 \\
        \textbf{HS/LS pairs}          & \textbf{14.92} & \textbf{12.38} & \textbf{14.10} & \textbf{0.95} & \textbf{0.90} & \textbf{0.90} \\
        \bottomrule
    \end{tabular}
\end{table*}

\begin{figure*}[p]
    \centering
    \includegraphics[width=120mm]{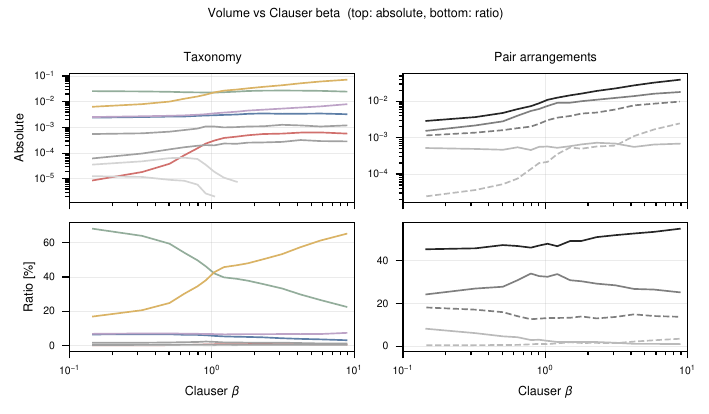}\\[6pt]
    \includegraphics[width=120mm]{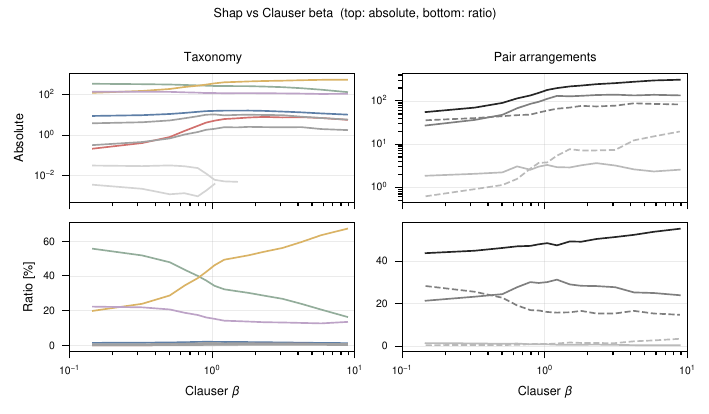}
    \caption{\textbf{Volume and importance budgets along the pressure-gradient ramp.} Top block, structure volume; bottom block, importance $\sum|\mathbf{\Phi}|$. Within each block the upper row gives the absolute value and the lower row the share of the total, with the classes of the main text's classification tree on the left (same colours) and the pair class split into its three arrangement axes on the right (solid black for all pooled spanwise; solid dark gray for streamwise with HS downstream, dashed for HS upstream; solid light gray for HS on top, dashed at bottom). The left share is of the total over all structures and the right share is of the pair class alone. Structures are binned by the chord position of their centroid and each bin is labelled by its Clauser $\beta$.}
    \label{sfig:census_vol_shap}
\end{figure*}

\begin{figure*}[p]
    \centering
    \includegraphics[width=120mm]{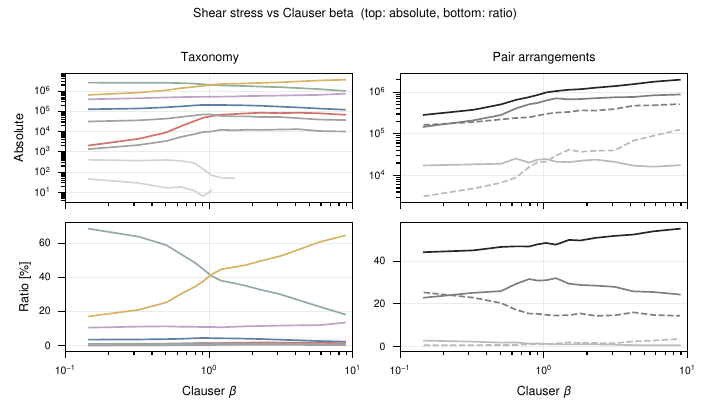}\\[6pt]
    \includegraphics[width=120mm]{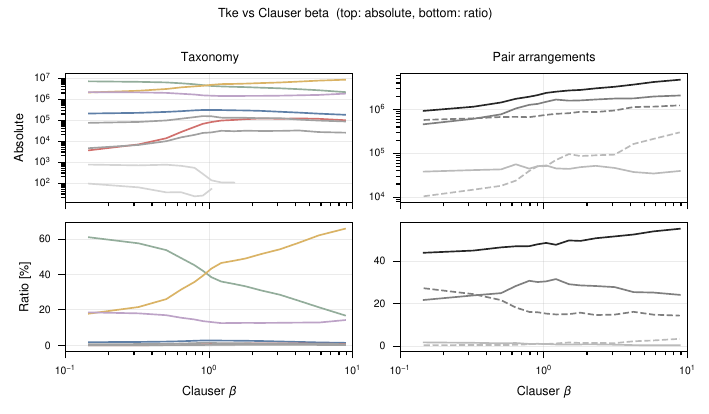}
    \caption{\textbf{Stress and energy budgets along the pressure-gradient ramp.} Reynolds shear stress $\sum|uv|$ (top block) and the turbulent kinetic energy $\sum k$ (bottom block), in the layout of Supplementary Fig.~\ref{sfig:census_vol_shap}.}
    \label{sfig:census_uv_k}
\end{figure*}

\paragraph{The census along the chord.} Supplementary Figs.~\ref{sfig:census_vol_shap} and \ref{sfig:census_uv_k} resolve the same five budgets against $\beta$, each figure carrying the absolute value on top and the share below, with the classes of the main text's classification tree on the left and the pair class split by arrangement on the right.

The absolute curves settle a question the shares leave open. The share of the volume held by pure LS structures falls from $68\%$ to $23\%$ along the ramp, which reads like a collapse. The absolute curve shows it is not: the volume of pure LS fluid is flat across two orders of magnitude of $\beta$. What changes is everything around it. The pair volume rises by more than an order of magnitude over the same range, and the pure HS volume roughly triples. The streak-to-pair takeover of the main text is therefore an addition and not a substitution: deceleration does not destroy the near-wall high- or low-speed structures; the momentum contrasts that surround them grow until they hold most of the flagged fluid. The importance, stress and energy budgets follow the volume closely, each crossing over between the single-speed structures and the pairs at $\beta\approx1$, so at the level of whole classes the importance budget tracks the volume budget. The deviations from that proportionality are what the density census of Supplementary Fig.~\ref{sfig:census_shapdens_uvdens} and the within-object asymmetry of Supplementary Note~6 measure.

Supplementary Fig.~\ref{sfig:census_shapdens_uvdens} resolves the density budgets. Here the absolute panels carry the physics and the share panels only rank. In absolute terms every class loses importance density along the ramp, by a factor of between 2.5 and 4 between $\beta\approx0.1$ and $\beta\approx10$. This is the same de-concentration the composites report: as the interface network expands, no single member of it stands out as sharply above its background. The shares, however, are almost flat. Pure HS fluid holds the largest share of the importance density at every station, ahead of pure Q4, and the pairs stay near the population mean throughout. The inversion between total and density is therefore not an artefact of pooling stations with different scalings; it holds at every pressure-gradient strength. The share panels of that figure each carry their own denominator, for the reason given above: the left column normalises over the nine classes it draws and the right column over the three pair arrangement axes alone. Their levels therefore differ from those of Supplementary Table~\ref{stab:census_density} and of main-text Fig.~1c, which use the seven pooled classes, and the three should be compared by ranking and by trend rather than value by value.

\begin{figure*}[p]
    \centering
    \includegraphics[width=120mm]{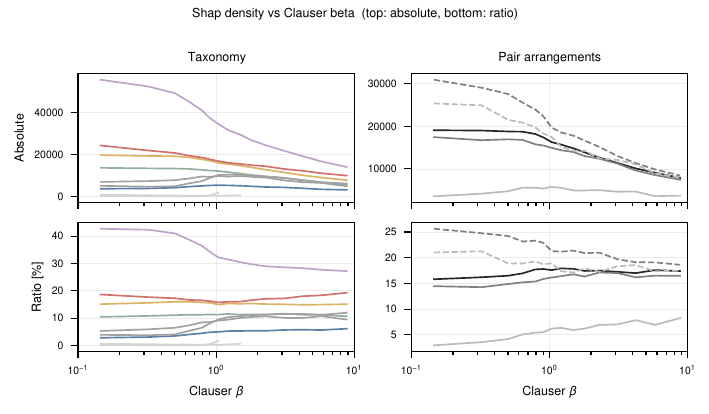}\\[6pt]
    \includegraphics[width=120mm]{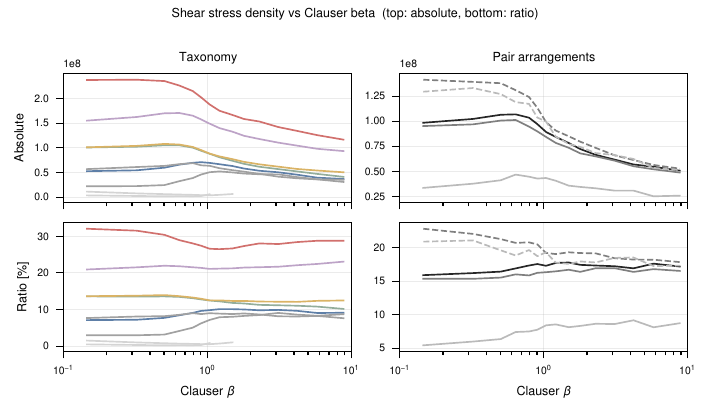}
    \caption{\textbf{Importance and stress densities along the ramp.} Importance per unit volume (top block) and Reynolds shear stress per unit volume (bottom block), absolute above and share below, in the layout of Supplementary Fig.~\ref{sfig:census_vol_shap}. The share panels normalise over the classes drawn in their own column, nine on the left and three on the right, so they reproduce the ranking of Supplementary Table~\ref{stab:census_density} and not its levels.}
    \label{sfig:census_shapdens_uvdens}
\end{figure*}

\clearpage
\paragraph{The census of the pair arrangements.} Supplementary Table~\ref{stab:census_pairs} gives the split of the pair class by the arrangement of its two lobes, which is the $55:38:5$ ranking of the main text in full. The volume-weighted and count-weighted views disagree, and the disagreement is informative rather than a nuisance. By count, spanwise and streamwise pairs are almost exactly as numerous as each other, $48\%$ against $47\%$, and the stacks are rare at $2\%$. By volume, the spanwise pairs take $54.5\%$ against $38.3\%$ for the streamwise ones, so the spanwise pairs are individually the larger objects. The main text quotes the volume-weighted split, because every other census in the study is volume-weighted, and notes the count-weighted parity.

Two entries deserve separate comment. The stacks are the only arrangement in which the two orientations are strongly unequal. Stacks with the fast lobe on top, the ordering of an undisturbed boundary layer, outnumber inverted stacks by $2.7$ to $1$ by count, yet the inverted ones hold more than twice the volume, $3.8\%$ against $1.5\%$ of the pair volume. An inverted stack is fast fluid trapped beneath slow fluid, a configuration that appears not to have been catalogued, and the census says it is rare but large. The two spanwise orientations, in contrast, are equal to within a percent, which is the symmetry the flow must obey.

\begin{table*}[ht!]
    \centering
    \caption{\textbf{The census of the pair arrangements.} Each pair is typed from the offset between the centroid of its low-speed lobe and that of its high-speed lobe, using $30^\circ$ half-angle cones about the wall-normal direction (stacks) and, for the wall-parallel remainder, about the streamwise direction (streamwise against spanwise), with the separation restricted to 5--400 wall units. Pairs whose lobes fall outside those windows are untyped. Shares are given as a percentage of the pair class.}
    \label{stab:census_pairs}
    \small
    \vspace{4pt}
    \begin{tabular}{lrrr}
        \toprule
        arrangement & count (\%) & volume (\%) & of which, by volume \\
        \midrule
        spanwise pair    & 48.3 & 54.5 & fast lobe right 26.4, left 28.1 \\
        streamwise pair  & 46.9 & 38.3 & fast lobe downstream 24.5, upstream 13.8 \\
        vertical stack   &  2.4 &  5.3 & fast lobe on top 1.5, below 3.8 \\
        untyped          &  2.4 &  1.9 & \\
        \bottomrule
    \end{tabular}
\end{table*}

\clearpage
% =====================================================================================
\subsection{Supplementary Note 4: where the two lobes of a pair sit}
% =====================================================================================

The arrangement census of Supplementary Table~\ref{stab:census_pairs} reduces each pair to one of three labels. This note gives the underlying distributions, which show that the labels report a real structure in the data rather than imposing one, and it reports how the separation between arrangements evolves.

\begin{figure*}[ht!]
    \centering
    \includegraphics[width=\linewidth]{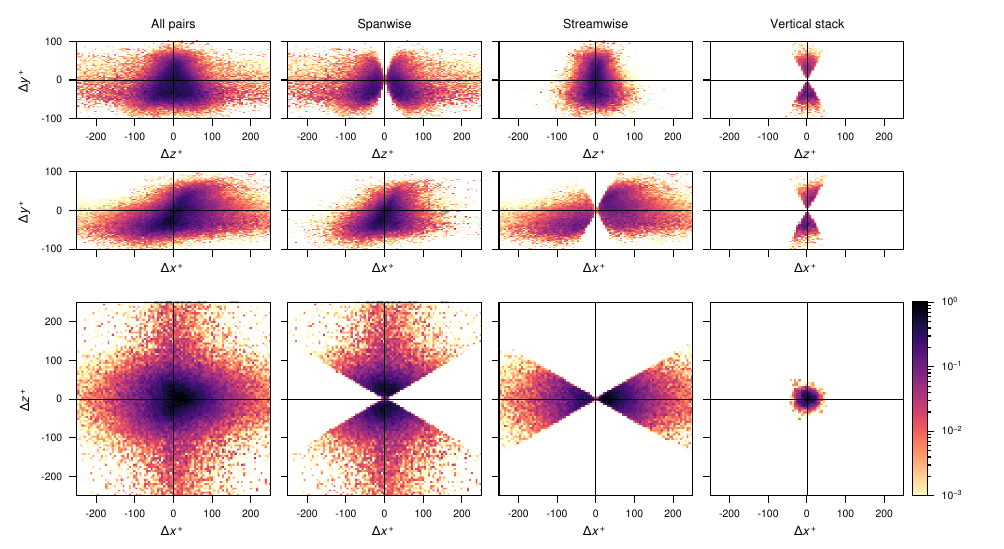}
    \caption{\textbf{The offset between the two lobes of a pair, projected on the three coordinate planes.} Volume-weighted joint density of the displacement from the low-speed lobe centroid to the high-speed lobe centroid, in wall units, for all pairs and for each pair arrangement (columns). Top, the cross-flow plane $(\Delta z^+,\Delta y^+)$; middle, the streamwise--wall-normal plane $(\Delta x^+,\Delta y^+)$; bottom, the wall-parallel plane $(\Delta x^+,\Delta z^+)$. Colour, density normalised to the maximum among all panels, on a logarithmic scale. The three arrangements of Supplementary Table~\ref{stab:census_pairs} are a partition of this distribution.}
    \label{sfig:jpdfs}
\end{figure*}

\paragraph{The offset distributions.} Supplementary Fig.~\ref{sfig:jpdfs} shows the joint distribution of the offset from the low-speed to the high-speed lobe, projected on the three coordinate planes. The wall-parallel projection (bottom row) is the one that carries the main-text claim. It is not isotropic: the density is elongated along the spanwise axis at small streamwise separation, so the most common way this flow binds fast fluid to slow fluid is side by side across the span. The streamwise arm is longer but thinner, which is the geometric statement behind the count-against-volume disagreement of Supplementary Table~\ref{stab:census_pairs}. The cross-flow projection (top row) shows the same anisotropy from the other side, with a compact spanwise core and only a thin vertical tail, the tail being the stacks. The distributions are smooth and single-peaked, with no gap at the cone boundaries, which is why the three arrangements should be read as the three principal directions of one continuous distribution rather than as three separate species.

\begin{figure*}[ht!]
    \centering
    \includegraphics[width=\linewidth]{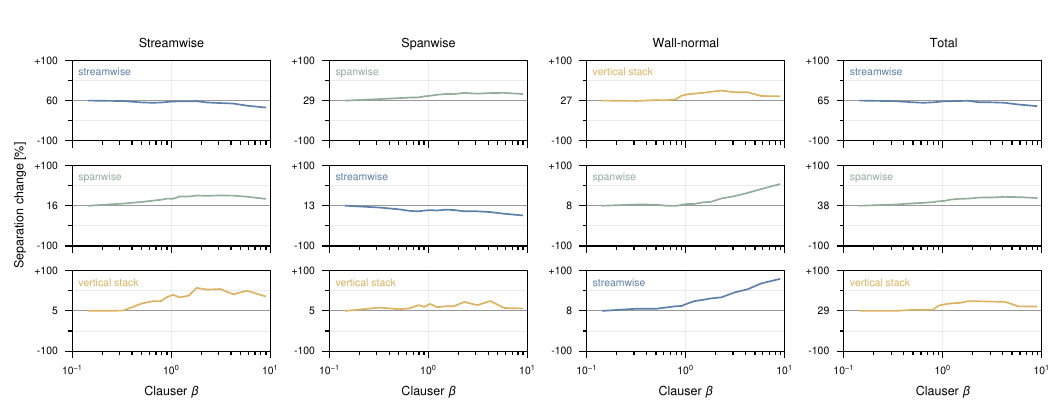}
    \caption{\textbf{The separation between the two lobes of a pair along the pressure-gradient ramp.} Median change of the separation between the low-speed and the high-speed lobe centroids, with respect to its value at the first station and in local wall units, against Clauser $\beta$, for the three arrangements. Panels, left to right: the streamwise, spanwise and wall-normal components, and the magnitude of the offset.}
    \label{sfig:sep}
\end{figure*}

\paragraph{The separations are compact and frozen in wall units.} Supplementary Fig.~\ref{sfig:sep} gives the median separation between the lobes, resolved into its magnitude and its three components, against $\beta$. Each arrangement is compact in its own direction, and every curve is flat or nearly so along the chord. The spanwise pairs separate by about $31$ wall units across the span and by about $18$ along the flow; the streamwise pairs by about $57$ wall units along the flow; the stacks by about $29$ wall units in height. These numbers do not drift while $\delta_{99}$ grows and $u_\tau$ falls by a factor of five. That is the geometric counterpart of the frozen composites of Supplementary Fig.~\ref{sfig:comp_pairs}: the pairs keep their viscous-unit size while their velocity contrast grows, which is the frozen viscous-unit geometry the main text reports.

\clearpage
% =====================================================================================
\subsection{Supplementary Note 5: the composite atlas}
% =====================================================================================

Main-text Fig.~3 shows two conditional-average fields ($u^+$ and $v^+$) for four of the arrangements, which is what the argument there needs. This note gives the complete atlas: every field for every arrangement and for every single-lobed class, the interface-normal profiles and the numbers measured from them, the three-dimensional shape of the composites, and their response to the pressure gradient. Conditional averages are ensemble averages over all instances of a population, sampled on a fixed grid of wall-unit offsets centred on the reference point of each object, so random turbulence cancels and only what the instances share survives.

\begin{figure*}[p]
    \centering
    \includegraphics[width=\linewidth]{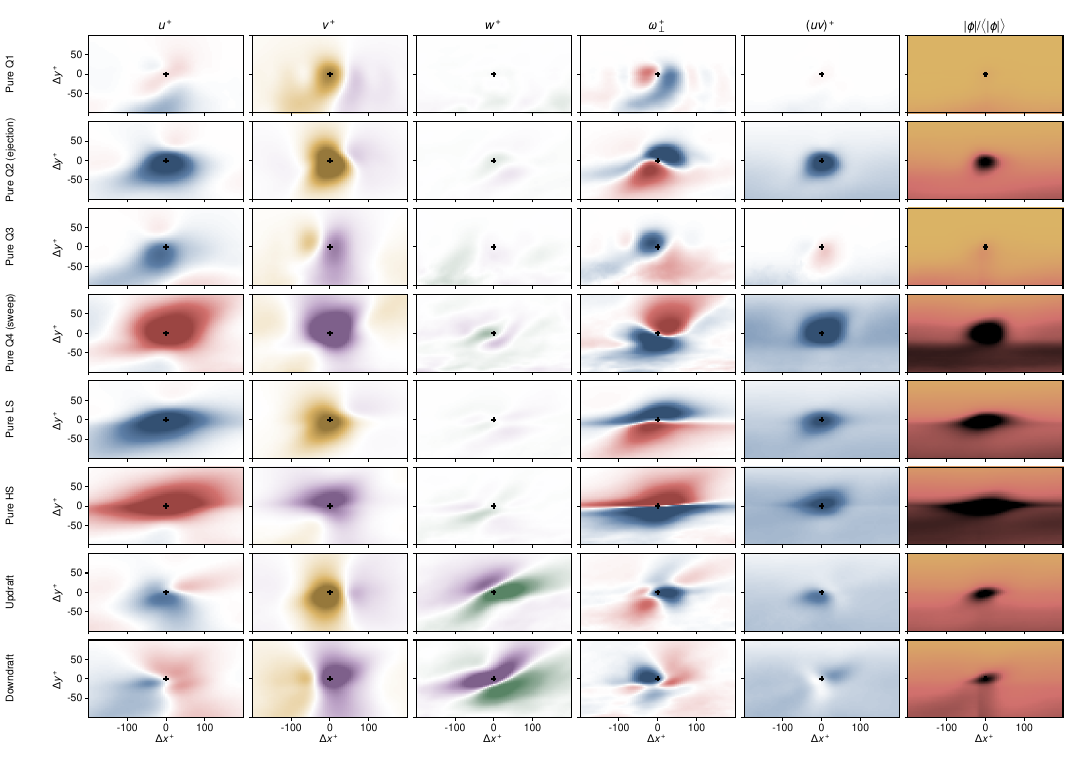}
    \caption{\textbf{The single-lobed classes, in every conditional-averaged field.} Rows, top to bottom: the pure Q classes (Q1, Q2, Q3 and Q4), pure LS structures, pure HS structures, updrafts and downdrafts. Columns, left to right: streamwise velocity $u^+$ (blue to red, limits $\mp2.0$), wall-normal velocity $v^+$ (purple to gold, limits $\mp0.9$), spanwise velocity $w^+$ (green to purple, limits $\mp0.4$), the in-plane vorticity (blue to red, limits $\mp0.06$), the Reynolds shear stress $(uv)^+$ (blue to red, limits $\mp4.6$), and the importance magnitude relative to its snapshot mean (orange to black, from 0 to 3.8). The composites are centred on the structure centroid (cross marker) and shown in the streamwise--wall-normal plane.}
    \label{sfig:comp_pure}
\end{figure*}

\paragraph{The single-lobed classes and pair arrangements, in every field.} Supplementary Fig.~\ref{sfig:comp_pure} shows, for the pure motion types, all five fields: streamwise, wall-normal and spanwise velocity, the Reynolds shear stress, the in-plane vorticity, and the importance. These composites are shown because they are the control: if the interface picture captured by pairs were an artefact of centring the average on the midpoint between two lobes, it would appear here too.

\begin{figure*}[p]
    \centering
    \includegraphics[width=\linewidth]{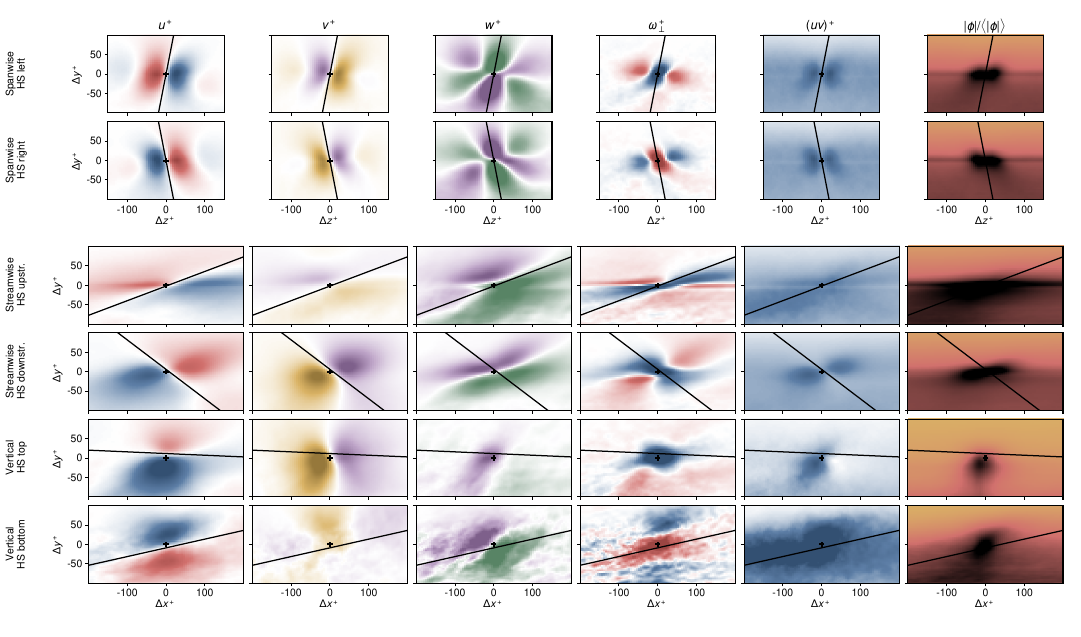}
    \caption{\textbf{The six pair arrangements, in every conditional-averaged field.} Rows, top to bottom: spanwise pairs with the fast lobe at positive and at negative spanwise offset, streamwise pairs with the fast lobe downstream and upstream, and vertical stacks with the fast lobe on top and below. Each row is shown on the plane holding its own momentum contrast, the cross-flow plane $(\Delta z^+,\Delta y^+)$ for the spanwise pairs and the streamwise--wall-normal plane $(\Delta x^+,\Delta y^+)$ for the others. Columns and colours as in Supplementary Fig.~\ref{sfig:comp_pure}. Black line, the fitted $u=0$ interface; cross, the composite centre.}
    \label{sfig:comp_pairs}
\end{figure*}

Supplementary Fig.~\ref{sfig:comp_pairs} gives the six pair arrangements with the same five fields. Each arrangement is shown on the plane that contains its own momentum contrast.

Reading across the rows makes the common rule visible. In every arrangement the two velocity lobes are separated by a thin $u=0$ surface, that surface carries a concentrated sheet of the vorticity component normal to the plane, and the importance is a compact region sitting on that surface. The differences between the rows are in orientation and strength, not in kind. The spanwise pair carries the near-vertical spanwise interface with its sheet of streamwise vorticity, and its spanwise velocity field is organised about the interface in the sense of that sheet, which is what a streamwise roll produces. The streamwise pair with its fast lobe downstream carries the inclined shear layer with its sheet of spanwise vorticity. The reversed streamwise pair carries the much thinner and flatter burst front. The stacks carry an almost horizontal interface, across which the wall-normal velocity forms an up- and downwash couple along the stream rather than a lobe-wise exchange. That is why the stack behaves as a boundary the flow rides over and not as a momentum pump.

\paragraph{The interface, measured.} For each composite the $u=0$ line is fitted inside a window around the centre, the fields are sampled along the normal to that line, and the velocity jump, the thickness, the vorticity-sheet amplitude and the importance contrast are read off the resulting profiles. As main-text Fig.~3 shows, the importance peak and the Reynolds-shear-stress peak are out of phase. The importance rises to a single maximum on the interface, while the stress has a local minimum there and two maxima in the lobe interiors on either side. Whatever the model is responding to, it is not the instantaneous momentum flux, a point Supplementary Note~6 pursues object by object.

Supplementary Table~\ref{stab:iface_pairs} collects every interface number for every arrangement, on both measurement planes. Three entries carry the arguments of the main text. The spanwise interface leans $11^\circ$ off the wall-normal and carries a velocity jump of $\Delta u^+=3.8$ across $26$ wall units, against $\Delta u^+=1.6$ across $26$ wall units for the inclined shear layer at $-37^\circ$: at the same thickness the spanwise interface carries more than twice the contrast. The two spanwise orientations agree to within $1.5\%$ in jump and to within $0.1^\circ$ in angle, which is the mirror symmetry the flow must have. The importance contrast is $2.1$--$2.4$ for the four spanwise and streamwise arrangements, so the model attends to the interface with comparable strength whichever way that interface faces.

\begin{table*}[ht!]
    \centering
    \caption{\textbf{Interface metrics of every composite.} Measured on the pooled composite of each row, on both planes. The angle is measured from the axis the interface runs along, that is from the streamwise axis on the $xy$ plane and from the wall-normal axis on the $zy$ plane; $\Delta u^+$ is the $10$--$90\%$ jump along the interface normal and $\delta_s^+$ the thickness over which it is taken; $\delta_\omega^+$ is the vorticity thickness; the importance contrast $|\mathbf{\Phi}|/\langle|\mathbf{\Phi}|\rangle$ is the ratio of the peak importance on the interface to the background of the composite; the aspect ratio is that of the region enclosed by the $90$th percentile of the importance. $N$ is the number of averaged objects.}
    \label{stab:iface_pairs}
    \vspace{4pt}
    \small
    \begin{tabular}{llrrrrrrr}
        \toprule
        Arrangement & Plane & $N$ & Angle (deg) & $\Delta u^+$ & $\delta_s^+$ & $\delta_\omega^+$ & $|\mathbf{\Phi}|/\langle|\mathbf{\Phi}|\rangle$ & AR \\
        \midrule
        Spanwise, fast right   & $zy$ & 176\,484 & $-11.0$ & 3.79 & 26.5 & 26.5 & 2.13 & 1.92 \\
        Spanwise, fast left    & $zy$ & 169\,808 & $+11.0$ & 3.74 & 27.5 & 27.7 & 2.13 & 1.98 \\
        Streamwise, fast downstream & $xy$ & 262\,533 & $-37.2$ & 1.63 & 26.0 & 24.4 & 2.36 & 3.80 \\
        Streamwise, fast upstream   & $xy$ &  73\,496 & $+20.7$ & 1.15 & 11.5 & 10.8 & 2.39 & 4.95 \\
        Vertical stack, fast on top     & $xy$ &  12\,306 & $-2.5$  & 3.18 & 26.5 & 25.7 & 2.33 & 1.58 \\
        Vertical stack, fast below      & $xy$ &   4\,615 & $+12.7$ & 3.35 & 38.5 & 33.4 & 1.80 & 2.55 \\
        \bottomrule
    \end{tabular}
\end{table*}

\begin{figure*}[p]
    \centering
    \includegraphics[width=\linewidth]{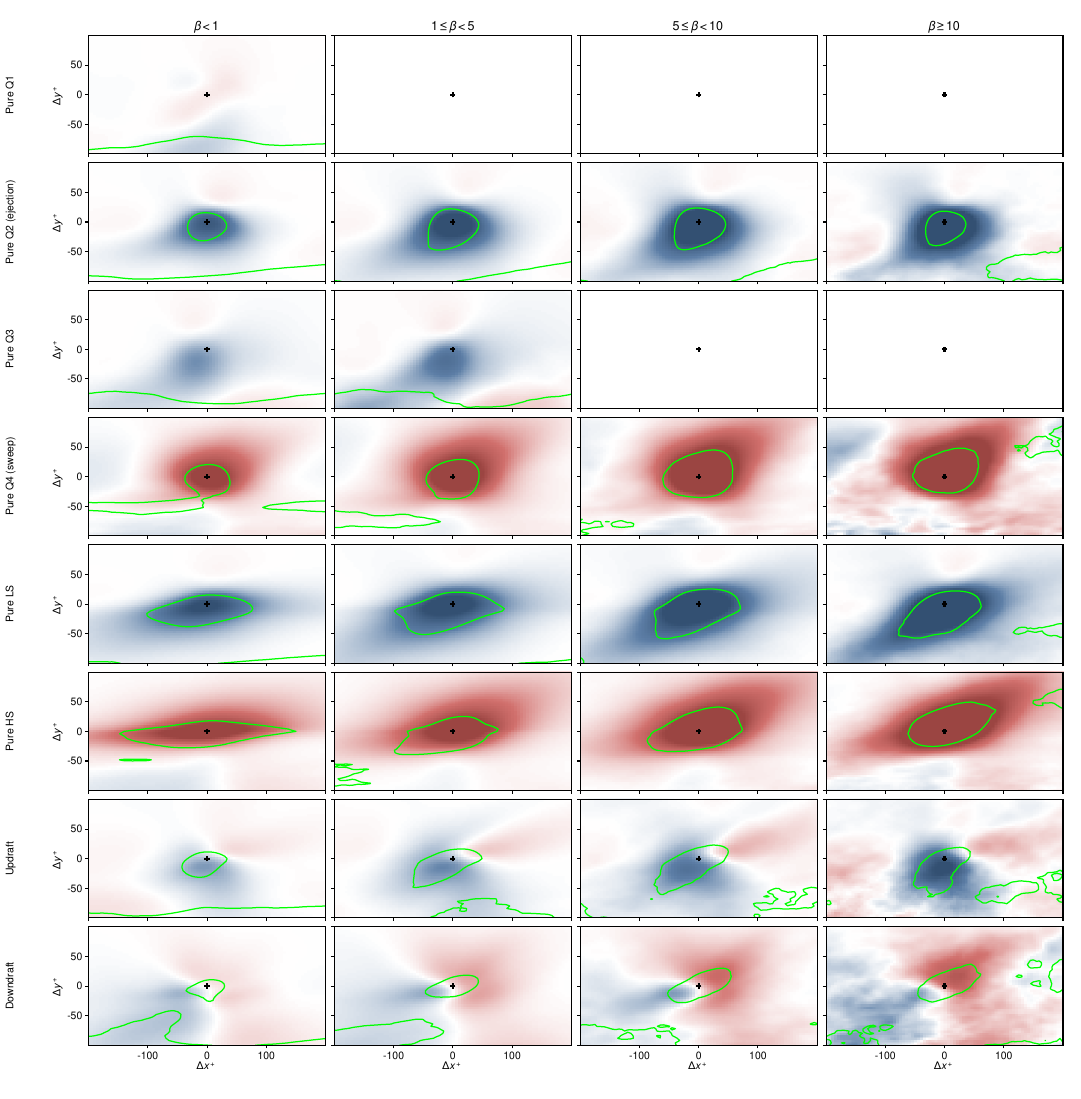}
    \caption{\textbf{The single-lobed composites rebuilt within bands of $\beta$.} Rows, the composites measured on the streamwise--wall-normal plane: the pure Q classes (Q1, Q2, Q3 and Q4), pure LS structures, pure HS structures, updrafts and downdrafts. Columns, four bands of Clauser $\beta$ ($\beta<1$, $1$--$5$, $5$--$10$, $\ge10$), Colour, streamwise velocity in viscous units on a scale shared by every panel (blue to red, limits $\mp2.85$); green contour, the $90$th percentile of the importance. Composites with less than 200 samples are left blank.}
    \label{sfig:si_comp_beta_pure}
\end{figure*}

\begin{figure*}[p]
    \centering
    \includegraphics[width=\linewidth]{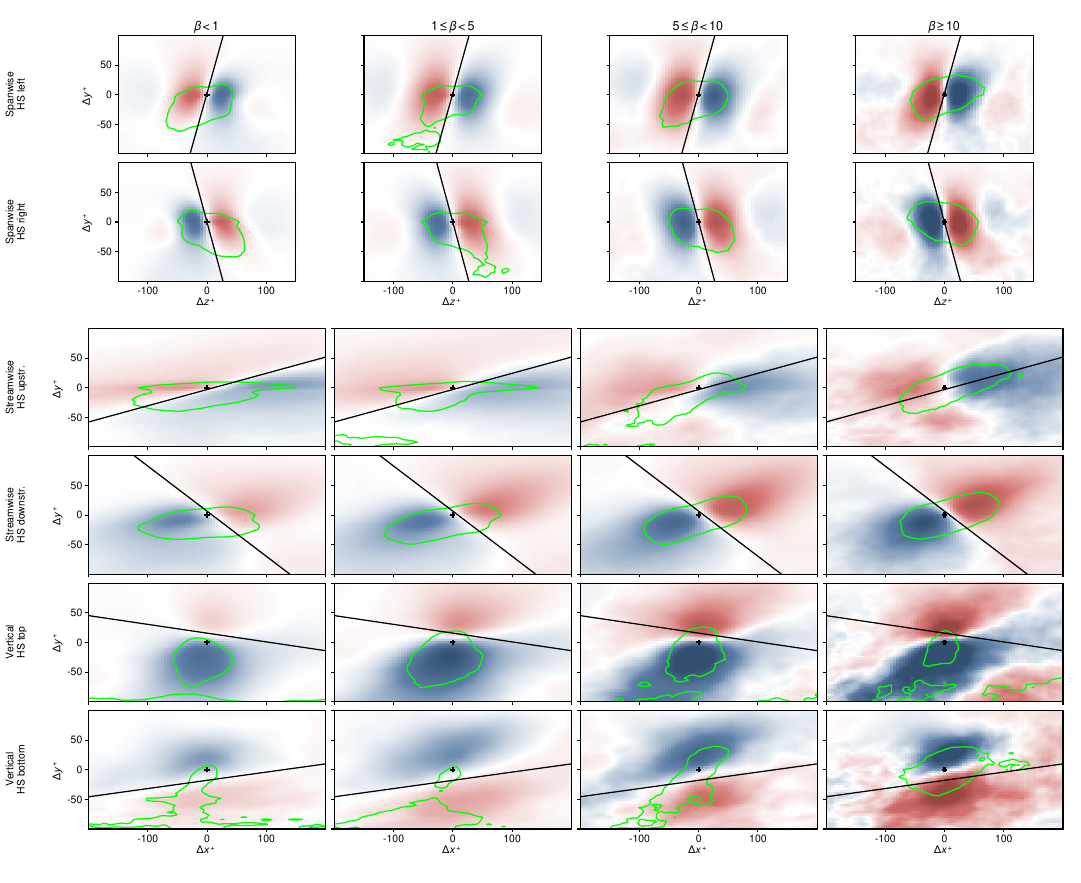}
    \caption{\textbf{The six pair-arrangement composites rebuilt within bands of $\beta$.} As Supplementary Fig.~\ref{sfig:si_comp_beta_pure}, for the composites measured on the crossflow plane for the spanwise pairs and the streamwise--wall-normal for the rest. The black line marks the fitted $u=0$ interface.}
    \label{sfig:si_comp_beta_pairs}
\end{figure*}

\paragraph{The response to the pressure gradient.} Supplementary Figs.~\ref{sfig:si_comp_beta_pure} and \ref{sfig:si_comp_beta_pairs} rebuild every composite within four bands of $\beta$. The result is the same for all of them: the geometry is frozen and only the amplitude changes. The inclined shear layer keeps its angle constant to within a couple of degrees and its thickness within $25$--$30$ wall units across the whole ramp. The spanwise interface keeps its thickness and steepens slightly towards vertical, from $18^\circ$ off the wall-normal in the weakest band to $7^\circ$ in the strongest. The stacks flatten. In all cases the velocity contrast between the lobes grows by a factor of two to three in friction units.

% Supplementary Figs.~19 and 20 turn those maps into curves against $\beta$, using the finer chord bins the trends can afford. Four of the six panels carry results used in the main text. The velocity jump grows for every arrangement, with the transverse interfaces keeping two to three times the jump of the inclined layer throughout. The thickness is flat for the well-populated arrangements, and the two panels in which it is not, the inverted stack and the near-wall front, are also the two rarest, where the composite is built from a few thousand objects and the interface fit becomes noisy. The interface angle is flat for every arrangement. And the importance contrast declines for every arrangement, by roughly a third between the mildest and the strongest band for the well-populated ones, which is the de-concentration reported in the main text: deceleration multiplies interfaces rather than sharpening the ones already there.

The growth of the friction-scaled jump is expected, because the friction velocity tends to zero as separation approaches, so friction scaling has to fail there by construction. Main-text Fig.~5c puts the wall-unit and the local-intensity scalings side by side for the four spanwise and streamwise arrangements. In wall units every jump grows steeply, and most steeply in the last band. Normalised by the local root-mean-square streamwise fluctuation $u'$, evaluated at the reference point of each object, the same curves flatten. For the inclined shear layer the normalised jump varies by about $\pm10\%$ over two decades of $\beta$, which is the flat curve of main-text Fig.~5c. The spanwise interface is stronger and its collapse is close but not exact, drifting from about $1.86$ to about $1.22$, so part of its strength is set by something other than the local turbulence intensity. The burst front, the weakest arrangement, falls from about $0.57$ to about $0.27$. The right-hand panel completes the picture from the attribution side. The aspect ratio of the region the model attends to is frozen at about $2$ for the spanwise pairs and lies between $3$ and $6$ for the streamwise ones, declining somewhat towards separation. That decline is coherent with the well-known shortening of the structures as APG increases.

\clearpage
% =====================================================================================
\subsection{Supplementary Note 6: importance is not instantaneous Reynolds stress}
% =====================================================================================

The most economical explanation of the census would be that the model simply flags fluid carrying high instantaneous Reynolds shear stress, in which case the importance field would be a restatement of quadrant analysis. The interface profiles of main-text Fig.~3 already argue against it, because the two fields peak in different places. This note gives the sharper test, which is made inside single objects and is therefore immune to any change of scale between stations.

Every pair contains a slow and a fast region. Summing a quantity over each region and forming the normalised difference gives an asymmetry that is $-1$ when the quantity lives entirely on the slow side, $0$ at perfect balance, and $+1$ on the fast side. For the Reynolds shear stress that asymmetry is $a_{uv} = \left(\sum|uv|_{u>0} - \sum|uv|_{u<0}\right)/\sum|uv|$, summed over the voxels of each region, and the importance asymmetry $a_\phi$ replaces $|uv|$ by $|\mathbf{\Phi}|$ in the same expression. Because it is a ratio within one object, it does not depend on the local friction velocity or on the size of the object, so it can be compared across the whole chord.

\begin{figure*}[ht!]
    \centering
    \includegraphics[width=60mm]{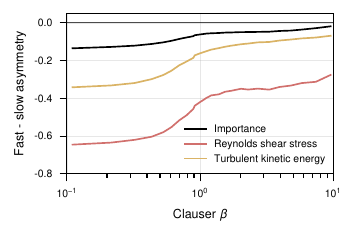}
    \caption{\textbf{Asymmetry of the Reynolds shear stress and of the importance within single structures.} Normalised asymmetry between the slow and fast regions of each structure, against Clauser $\beta$, for the Reynolds shear stress and for the importance. The asymmetry is $-1$ when the quantity lives entirely in the points of slow streamwise velocity ($u<0$), $0$ at balance, and is computed within each object, so it is independent of the local scaling.}
    \label{sfig:asymmetry}
\end{figure*}

Supplementary Fig.~\ref{sfig:asymmetry} gives that asymmetry for the Reynolds shear stress and for the importance. The stress asymmetry starts at $-0.61$ at $\beta\approx0.1$, so at weak APG the slow fluid of a pair carries about four times the stress of its fast counterpart, and it closes to $-0.20$ at $\beta\approx10$ as the sweeps strengthen. The importance asymmetry follows the same trend but stays far closer to balance throughout, from $-0.11$ to $-0.02$. At every station the model weighs the two parts of an object far more evenly than the physics distributes the stress between them, and by $\beta\approx10$ it weighs them almost equally.

Two conclusions follow. The importance field measures predictive relevance and not instantaneous momentum flux, which extends to a decelerating layer the observation, made in channel flow, that the most important structures are not those carrying the most stress. The practical corollary is that slow fluid is nearly as informative as fast fluid everywhere there is an interface, and increasingly so as separation is approached. That is a statement about where a sensor or an actuator should be placed.

\clearpage
% =====================================================================================
\subsection{Supplementary Note 7: the morphology of the importance structures}
% =====================================================================================

Main-text Fig.~2c compares the median elongation and inclination of the four families. This note gives the distributions behind those medians, adds the two aspect ratios the main figure has no room for, and reports where the structures sit relative to the wall.

As main-text Fig.~2c shows, at weak APG each family has its own signature: streaks are strongly elongated in the streamwise direction, with a median length-to-height ratio near $6.9$, importance structures are intermediate at $3.5$, and vortices and intense Q events are compact at $2.3$ and $1.8$. Deceleration erases those signatures. By $\beta\approx10$ all four families have shortened to a comparable and modest anisotropy, and all four have steepened, by between three and seven degrees. Geometry can no longer differentiate the families near separation, which is precisely why the main text turns to what the structures contain and how they are arranged.

\begin{figure*}[p]
    \centering
    \includegraphics[width=\linewidth]{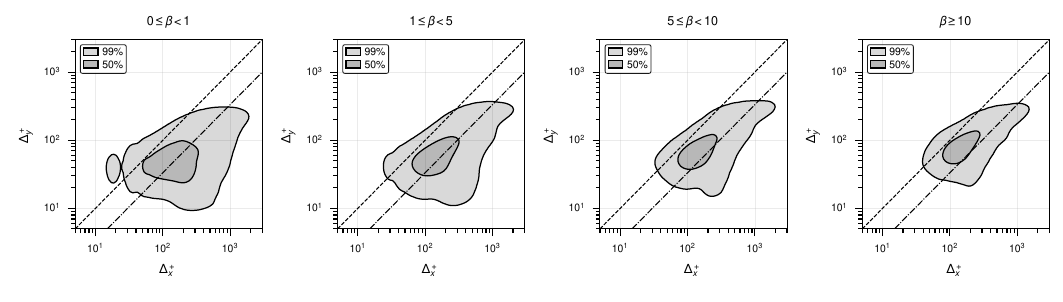}\\[6pt]
    \includegraphics[width=\linewidth]{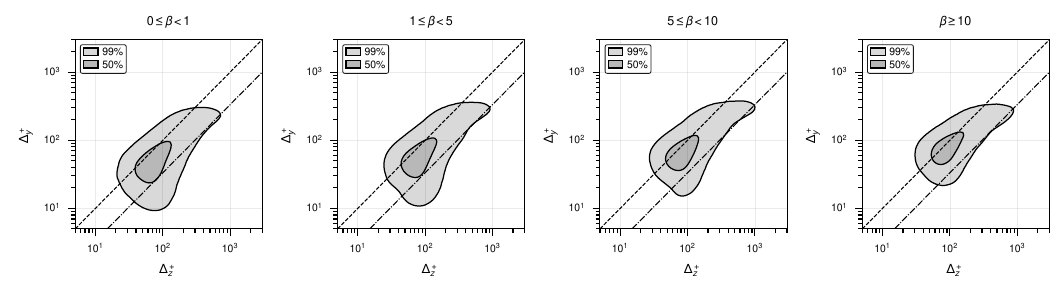}\\[6pt]
    \includegraphics[width=\linewidth]{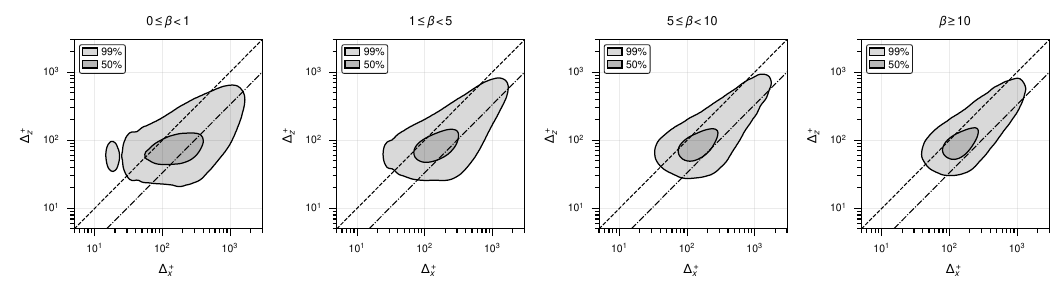}
    \caption{\textbf{Joint size distributions of the importance structures.} Bounding dimensions of each structure in wall units, on logarithmic axes, in four bands of Clauser $\beta$ (columns: $0\le\beta<1$, $1\le\beta<5$, $5\le\beta<10$, $\beta\ge10$). Top, streamwise length against wall-normal height; middle, spanwise width against wall-normal height; bottom, streamwise length against spanwise width. Contours enclose the $50\%$ and $99\%$ densest population; the dash-dotted diagonals are lines of constant aspect ratio.}
    \label{sfig:dims}
\end{figure*}

Supplementary Fig.~\ref{sfig:dims} gives the joint distributions of the bounding dimensions of the importance structures in the three coordinate planes resolved in four bands of $\beta$. The spanwise dimension behaves differently from the streamwise one. In the $\Delta z^+$--$\Delta y^+$ plane the contours sit close to the diagonal at every $\beta$, so the structures are roughly as wide as they are tall throughout, whereas in the $\Delta x^+$--$\Delta y^+$ plane they start well below the diagonal and move onto it. The shortening the main text reports is therefore a streamwise shortening, and it brings the structures towards the cross-sectional proportions they already had.

\clearpage
% =====================================================================================
\subsection{Supplementary Note 8: the comparison with the classical families}
% =====================================================================================

The central control of the study is the head-to-head comparison with the three classical families, detected from the same instantaneous fields with their own criteria and with thresholds fixed by the same percolation analysis. Those criteria are intense Q events at $|uv|>1.62\,u'v'$, streamwise streaks at $\sqrt{u^2+w^2}>3.81\,u_\tau$, and vortex clusters at $\Delta>0.114\,\Delta'$, against $|\mathbf{\Phi}|>H_\Phi^\star\,\Phi_{\rm rms}$ with $H_\Phi^\star=1.71$ for the importance structures. Main-text Fig.~2 reports the headline overlap and the morphology. This note gives the comparison in full, in three independent spaces: the velocity--wall-distance plane, the quadrant plane, and the voxel-wise overlap.

\begin{figure*}[p]
    \centering
    \includegraphics[width=\linewidth]{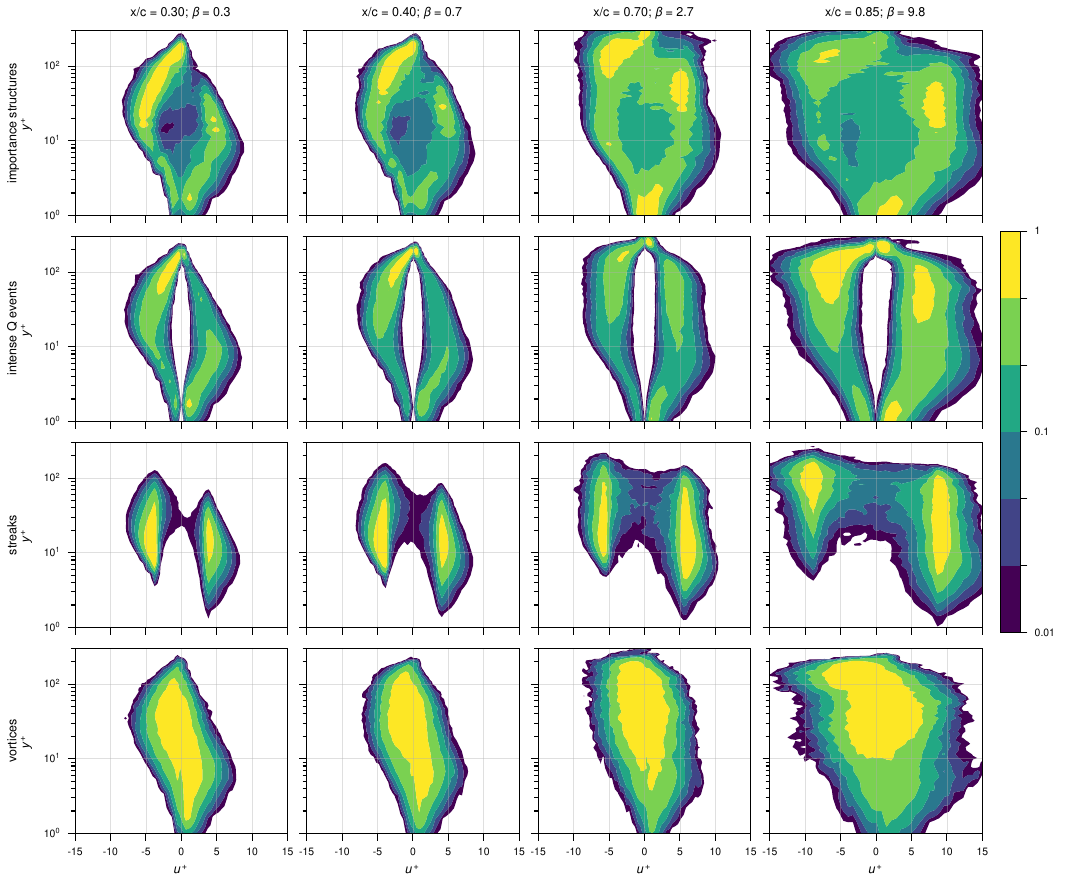}
    \caption{\textbf{The fluid each family selects, in viscous units.} Joint probability density function of the streamwise velocity $u^+$ and the wall distance $y^+$ of the flagged fluid, at four stations (columns) for the four structure families (rows): importance structures, intense Q events, streaks and vortices. Colour, joint density normalised to the maximum of its own panel, on a logarithmic scale.}
    \label{sfig:hist_plus}
\end{figure*}

\begin{figure*}[p]
    \centering
    \includegraphics[width=\linewidth]{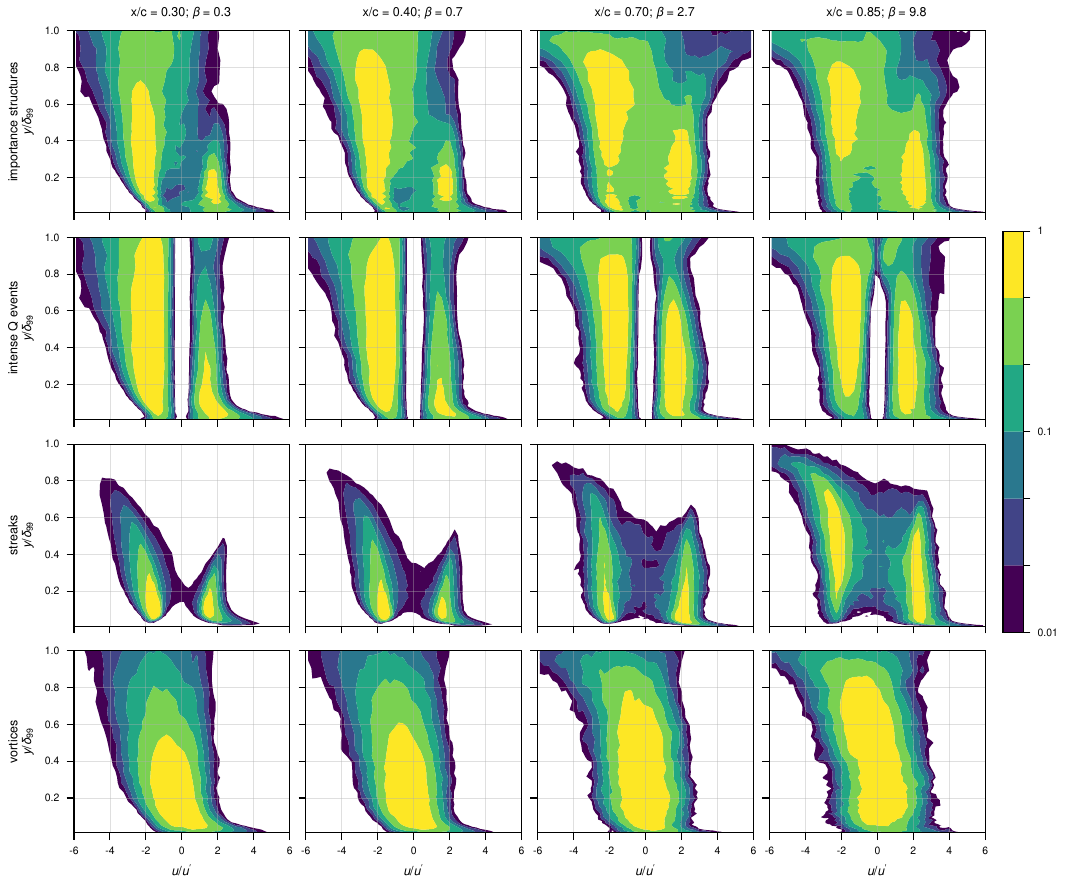}
    \caption{\textbf{The same comparison in local units.} As Supplementary Fig.~\ref{sfig:hist_plus}, with the streamwise velocity normalised by the local root-mean-square fluctuation $u'(y)$ of the same station and the wall distance by the local boundary-layer thickness, on a linear height axis cut at $y=\delta_{99}$. Above that height $u'$ decays into the free stream while $u$ does not, so the ratio stops being an amplitude and the panels end there.}
    \label{sfig:hist_outer}
\end{figure*}

\paragraph{The fluid each family selects.} Supplementary Figs.~\ref{sfig:hist_plus} and \ref{sfig:hist_outer} histogram the streamwise velocity and the wall distance of the fluid flagged by each family, at four stations, in two scalings. The first is viscous, $(u^+,y^+)$: the velocity is divided by the friction velocity and the height by the viscous length. The second is local, $(u/u',y/\delta_{99})$: the velocity is divided by the local root-mean-square fluctuation $u'(y)$ of the same station and the height by the boundary-layer thickness. Both are needed because neither survives the whole ramp on its own. The pressure gradient collapses $u_\tau$ while the turbulence intensity does not follow it, so the two scalings evolve in opposite directions along the chord. Only the pair of figures separates a change in the flow from a change in the units it is measured in.

In viscous units the four families separate cleanly. The intense Q events show the hyperbolic hole as a gap along $u^+=0$ that widens with wall distance, which is what their criterion imposes. The streaks are bimodal, two narrow lobes at $u^+\approx\pm4$ with nothing between them. The vortices fill a broad region centred on the mean velocity. The importance structures span the velocity range of the streak and Q-event fluid combined and, unlike either, keep the moderate-amplitude fluid between them: their distribution is continuous through $u^+=0$. Every distribution also widens from station to station, which reads as objects of growing amplitude. The local scaling shows that most of that growth is an artefact of viscous scaling, and that the genuine change is the growth of a high-speed peak near the wall.

The local scaling removes the $u_\tau$ dependency. Measured in $u'$, the four distributions barely change shape between $\beta=0.3$ and $\beta=9.8$: the importance peaks sit close to $u/u'=\pm2$ at every station, the intense Q events keep theirs at $\pm1.6$, and the hole stays open between them at every height and every station. Most of the widening seen in Supplementary Fig.~\ref{sfig:hist_plus} is therefore the collapse of $u_\tau$ and not a growth of the objects, which is the same conclusion the interface amplitudes reach in Supplementary Note~5. The hole is the one feature that survives every rescaling, because the criterion imposes it, and it remains the clearest statement of what intensity-based detection cannot see.

Three changes do survive the rescaling, quoted here as fractions of the flagged volume. The first is a move towards balance between the two signs of $u$. The low-speed share of the importance-flagged volume falls from $80\%$ at the weak-APG station to $58\%$ at the strong-APG station, and the Q-event fluid follows the same path more slowly, from $79\%$ to $62\%$. Neither crosses into high-speed dominance, but the important fluid approaches it faster than the intense quadrant fluid does, which is the pointwise counterpart of the within-object result of Supplementary Note~6. The second is the growth of the moderate fluid. The band $|u/u'|<1$ holds $16\%$ of the flagged volume at $\beta=0.3$ and $26\%$ at $\beta=9.8$, against $7$--$11\%$ of the Q-event volume at every station: the stronger the pressure gradient, the more of what the model flags is fluid of unremarkable amplitude. The third is the departure of the streaks from the wall. Half of the streak volume lies below $y=0.2\,\delta_{99}$ at the two weak-APG stations, against $18\%$ at the strong-APG station, where $35\%$ of it has moved above mid-layer. Its two lobes also separate in rms units, from $\pm1.6$ to $\pm2.2$. Part of this movement follows from the criterion rather than from the streaks. The streak threshold is a fixed multiple of $u_\tau$, one value per station with none of the wall-normal dependence the other three criteria carry. As $u_\tau$ collapses along the chord, the level at which streaks are cut therefore drifts relative to the local turbulence intensity. The departure from the wall and the widening of the two lobes should therefore be read as properties of the classical streak definition applied to a strongly decelerated layer, and not as an unconditioned statement about the flow. The importance-flagged fluid does none of this. About half of its volume sits above $y=0.5\,\delta_{99}$ at every station, so it is spread over the whole layer from the start. The vortices are the one family that moves without changing what it selects. Their fluid stays centred on the mean velocity, $57\%$ of its volume inside $|u/u'|<1$ at every station, while the share sitting above mid-layer grows from $29\%$ to $48\%$. Selecting mean-velocity fluid wherever that fluid sits is why their overlap with the importance set stays small.

\begin{figure*}[ht!]
    \centering
    \includegraphics[width=120mm]{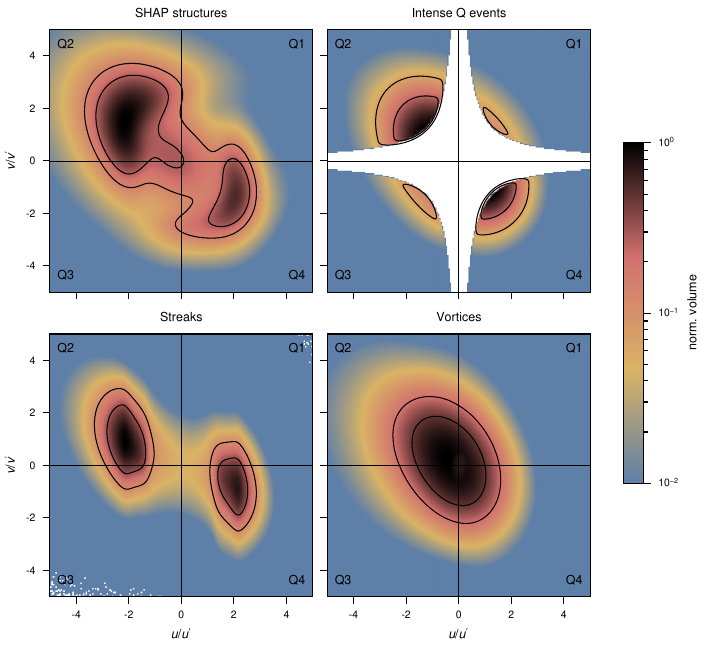}
    \caption{\textbf{The quadrant plane of every grid point inside each family.} Volume-weighted joint density of the local fluctuations $(u/u',v/v')$ over all points inside the structures of each family: importance structures (top left), intense Q events (top right), streaks (bottom left) and vortices (bottom right). Colour, density on a logarithmic scale, normalised to the maximum of each panel; solid contours enclose the $50\%$ and $75\%$ densest fluid.}
    \label{sfig:quad}
\end{figure*}

\paragraph{The quadrant plane, point by point.} Supplementary Fig.~\ref{sfig:quad} places every grid point inside a structure of each family in the quadrant plane of the local fluctuations. This is the pointwise counterpart of the structure-level planes of Supplementary Fig.~\ref{sfig:quadrant}, and it is the cleanest single statement of what distinguishes the importance field from the classical criteria.

The Q-event panel is empty inside the hyperbolic hole, by construction. The streak panel has two dense peaks at $u/u'\approx\pm2$ that reach across both signs of $v$, and its density drops by more than an order of magnitude between them. The vortex panel is broad, weakly organised, and centred on the origin, which says that vortex cores contain fluid of every kind and select none. The importance panel has the two sweep-ejection peaks of the Q panel, at almost the same positions, joined by a continuous band that runs straight through the hole. The importance field therefore contains the intense quadrant fluid, contains the streak fluid, and adds the connecting fluid that neither criterion retains. That connecting fluid is the interface fluid of the composites, and no intensity threshold can hold it together with the lobes it joins.

\begin{figure*}[p]
    \centering
    \begin{minipage}{0.46\linewidth}\begin{flushright}\includegraphics[width=60mm]{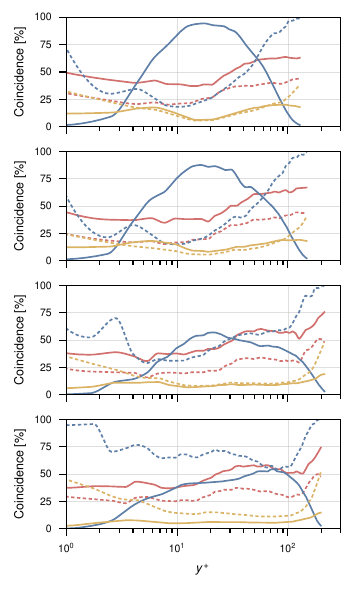}\end{flushright}\end{minipage}
    \begin{minipage}{0.46\linewidth}\begin{flushleft}\includegraphics[width=60mm]{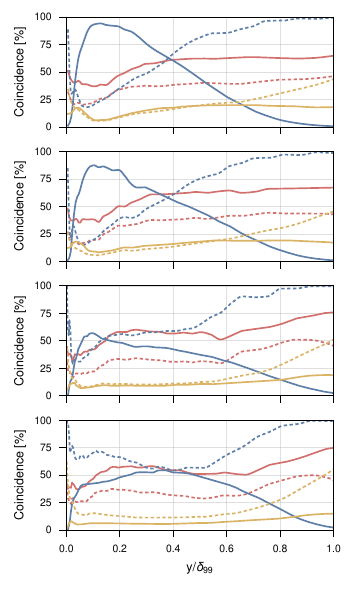}\end{flushleft}\end{minipage}\\[4pt]
    \caption{\textbf{Voxel overlap between the importance structures and the three classical families, at four stations.} Coincidence against wall distance in viscous units (left column) and in outer units (right column) at $x/c=0.30$, $0.40$, $0.70$ and $0.85$ from top to bottom. Solid, the composition of the flagged volume, $(X\cap S)/S$; dashed, the capture rate $(X\cap S)/X$; red, intense Q events; blue, streaks; gold, vortices. Main-text Fig.~2b shows the first and last panels of the left column.}
    \label{sfig:coinc}
\end{figure*}

\paragraph{The voxel overlap.} The third comparison uses two ratios. The composition of the importance volume is the fraction of it that belongs to a given classical family, and the capture is the fraction of that family flagged as important. Supplementary Fig.~\ref{sfig:coinc} gives both against wall distance at the four reference stations, in viscous and in outer units. Main-text Fig.~2b shows the first and the fourth station of the viscous-unit figure.

The trend along the chord is a dissolution. At $x/c=0.30$ the composition is a streak monopoly: in the buffer layer more than $90\%$ of the flagged fluid is streak fluid, the Q events contribute roughly $40$--$65\%$ at all heights, and the vortices never exceed $20\%$. By $x/c=0.85$ the streak peak has gone, the composition is spread across families and across heights, and the Q-event and streak contributions have become comparable at about half the flagged volume each. The capture rates sharpen this. The near-wall streaks that survive are almost entirely flagged as important, with capture rates above $90\%$ below $y^+\approx5$ at the strong-APG stations. The model has therefore not stopped attending to streaks; there are simply fewer of them where it looks. The Q-event capture stays moderate, between $25\%$ and $50\%$, at every station and every height. The vortex capture falls steadily along the chord, to about $10\%$ at the strong-APG stations: vortices are the least relevant of the three families, and deceleration weakens their relevance further. The rise of the dashed curves towards $100\%$ at the outermost heights is geometric, not physical, and reflects the vanishing volume of the classical families near the boundary-layer edge.

The outer-unit version shows that the dissolution is not a wall-scaling effect. Plotted against $y/\delta_{99}$ the streak composition peak at the weak-APG stations sits at $y/\delta_{99}\approx0.1$ and simply disappears downstream rather than moving, and the vortex capture decays at every height at once.

\begin{figure*}[p]
    \centering
    \includegraphics[width=\linewidth]{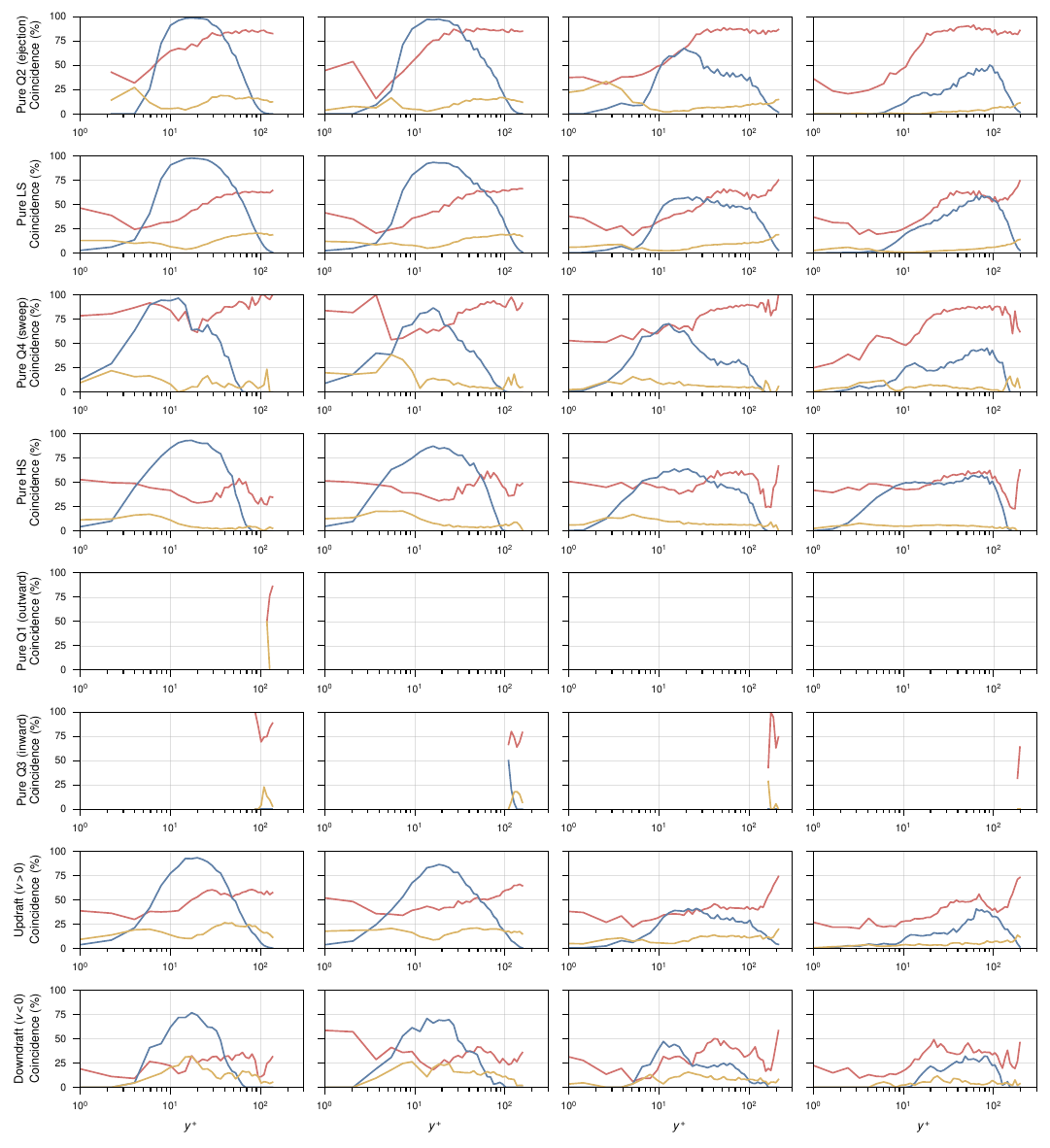}
    \caption{\textbf{Which classical structure each class of the taxonomy sits on.} Fraction of the volume of each taxonomy class that coincides with a classical structure, against wall distance in viscous units, at $x/c=0.30$, $0.40$, $0.70$ and $0.85$ (columns from left to right). Rows, the eight non-pair classes. Red, intense Q events; blue, streaks; gold, vortices. Curves are drawn only where the class has enough volume at that height, which is why the rare outward-inward rows are nearly empty.}
    \label{sfig:coinc_pure}
\end{figure*}

\begin{figure*}[p]
    \centering
    \includegraphics[width=\linewidth]{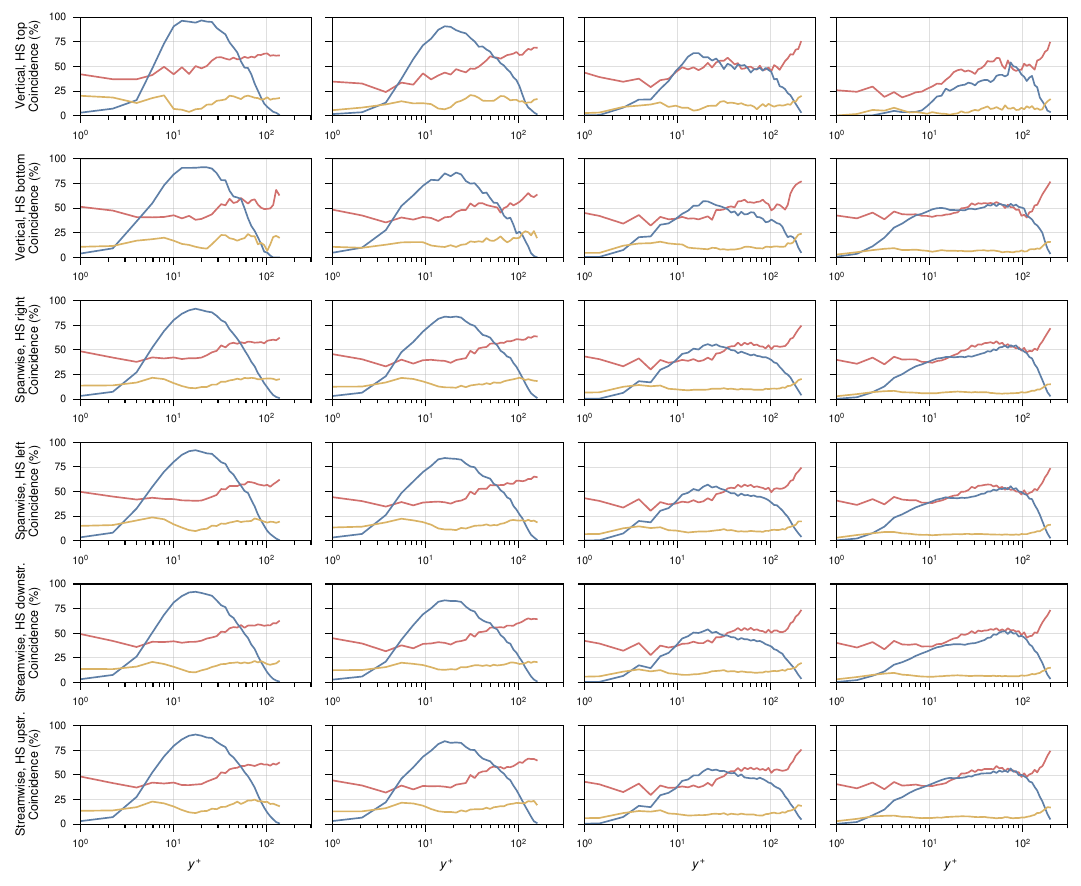}
    \caption{\textbf{The pair arrangements against the classical families.} As Supplementary Fig.~\ref{sfig:coinc_pure}, for the six pair arrangements of Supplementary Table~\ref{stab:census_pairs}.}
    \label{sfig:coinc_pairs}
\end{figure*}

\paragraph{Which classical structure each class of the taxonomy sits on.} The overlap can also be resolved by class, which turns it from a global control into a validation of the taxonomy itself. Supplementary Figs.~\ref{sfig:coinc_pure} and \ref{sfig:coinc_pairs} give the composition profiles for every class and for every pair arrangement.

The classes behave exactly as their names require, which is the strongest independent check of the classification available. The pure Q2 and pure Q4 classes sit on Q events over $81$--$86\%$ of their volume at every station, even though the classification never consults the Q-event segmentation and reads only the velocity content of each structure. The pure HS and pure LS structures sit on streaks over $74\%$ and $46\%$ of their volume at the weak-APG station, falling to $46\%$ and $33\%$ near separation as the streaks themselves become rarer. The pair arrangements are the informative entry. Every one of them sits about half on Q events and a third to a half on streaks, at every station, with no arrangement standing out. A pair is therefore not a relabelled Q event and not a relabelled streak; it is an object that contains part of both plus fluid that belongs to neither. The vortex overlap is small for every class and falls monotonically along the chord, from $7$--$19\%$ at the weak-APG station to $0$--$6\%$ at the strong-APG station, without exception.

%% file: references.bib
@book{pope_turbulent_2000,
	title = {Turbulent {Flows}},
	isbn = {978-0-521-59886-6},
	language = {en},
	publisher = {Cambridge University Press},
	author = {Pope, Stephen B.},
	year = {2000},
}

@book{anderson_fundamentals_2017,
	address = {New York},
	edition = {6},
	title = {Fundamentals of {Aerodynamics}},
	isbn = {978-1-259-25134-4},
	publisher = {McGraw-Hill Education},
	author = {Anderson, John D.},
	year = {2017},
}

@article{molina-casino_identification_2026,
	title = {Identification of most important structures in turbulent wings through explainable deep learning},
	volume = {3230},
	url = {https://doi.org/10.1088/1742-6596/3230/1/012029},
	doi = {10.1088/1742-6596/3230/1/012029},
	number = {1},
	journal = {Journal of Physics: Conference Series},
	publisher = {IOP Publishing},
	author = {Molina-Casino, Samuel and Cremades, Andrés and Hoyas, Sergio and Cardesa, José I. and Chedevergne, François and Vinuesa, Ricardo},
	month = may,
	year = {2026},
	pages = {012029},
}

@techreport{molina-casino_uncovering_2026,
	address = {Orlando, Florida, USA},
	title = {Uncovering {Dynamically} {Significant} {Coherent} {Structures} in {Wing} {Turbulence} through {Explainable} {Deep} {Learning}},
	doi = {10.2514/6.2026-0090},
	url = {https://arc.aiaa.org/doi/abs/10.2514/6.2026-0090},
	institution = {American Institute of Aeronautics and Astronautics},
	number = {AIAA 2026-0090},
	note = {AIAA SCITECH 2026 Forum},
	language = {en},
	urldate = {2026-01-19},
	author = {Molina-Casino, Samuel and Cremades, Andrés and Hoyas, Sergio and Cardesa, José I. and Chedevergne, François and Vinuesa, Ricardo},
	month = jan,
	year = {2026},
}

@article{chen_uniform-momentum_2021,
	title = {The uniform-momentum zones and internal shear layers in turbulent pipe flows at {Reynolds} numbers up to {Re}\_tau=1000},
	volume = {90},
	issn = {0142-727X},
	url = {https://www.sciencedirect.com/science/article/pii/S0142727X21000473},
	doi = {10.1016/j.ijheatfluidflow.2021.108817},
	urldate = {2026-07-21},
	journal = {International Journal of Heat and Fluid Flow},
	author = {Chen, Xue and Chung, Yongmann M. and Wan, Minping},
	month = aug,
	year = {2021},
	pages = {108817},
}

@article{montemuro_self-sustaining_2020,
	title = {A self-sustaining process theory for uniform momentum zones and internal shear layers in high {Reynolds} number shear flows},
	volume = {901},
	issn = {0022-1120, 1469-7645},
	url = {https://www.cambridge.org/core/journals/journal-of-fluid-mechanics/article/selfsustaining-process-theory-for-uniform-momentum-zones-and-internal-shear-layers-in-high-reynolds-number-shear-flows/81AF163DD000DCD612E528A3E8A4B94C},
	doi = {10.1017/jfm.2020.517},
	language = {en},
	urldate = {2026-07-21},
	journal = {Journal of Fluid Mechanics},
	author = {Montemuro, Brandon and White, Christopher M. and Klewicki, Joseph C. and Chini, Gregory P.},
	month = oct,
	year = {2020},
	pages = {A28},
}

@article{dong_coherent_2017,
	title = {Coherent structures in statistically stationary homogeneous shear turbulence},
	volume = {816},
	issn = {0022-1120, 1469-7645},
	url = {https://www.cambridge.org/core/journals/journal-of-fluid-mechanics/article/coherent-structures-in-statistically-stationary-homogeneous-shear-turbulence/E2A28C88B6D4F5925EBD0541A841D2FA},
	doi = {10.1017/jfm.2017.78},
	language = {en},
	urldate = {2026-07-21},
	journal = {Journal of Fluid Mechanics},
	author = {Dong, Siwei and Lozano-Durán, Adrián and Sekimoto, Atsushi and Jiménez, Javier},
	month = apr,
	year = {2017},
	pages = {167--208},
}

@article{gul_internal_2020,
	title = {Internal shear layers and edges of uniform momentum zones in a turbulent pipe flow},
	volume = {901},
	issn = {0022-1120, 1469-7645},
	url = {https://www.cambridge.org/core/journals/journal-of-fluid-mechanics/article/internal-shear-layers-and-edges-of-uniform-momentum-zones-in-a-turbulent-pipe-flow/EB19AC47ADBA8F3C975FC1D6A705B5E1},
	doi = {10.1017/jfm.2020.503},
	language = {en},
	urldate = {2026-07-21},
	journal = {Journal of Fluid Mechanics},
	author = {Gul, M. and Elsinga, G. E. and Westerweel, J.},
	month = oct,
	year = {2020},
	pages = {A10},
}

@article{fan_detection_2019,
	title = {On the detection of internal interfacial layers in turbulent flows},
	volume = {872},
	issn = {0022-1120, 1469-7645},
	url = {https://www.cambridge.org/core/journals/journal-of-fluid-mechanics/article/on-the-detection-of-internal-interfacial-layers-in-turbulent-flows/1600E66E62C06B467249C2A8E53B7E2E},
	doi = {10.1017/jfm.2019.343},
	language = {en},
	urldate = {2026-07-21},
	journal = {Journal of Fluid Mechanics},
	author = {Fan, Duosi and Xu, Jinglei and Yao, Matthew X. and Hickey, Jean-Pierre},
	month = aug,
	year = {2019},
	pages = {198--217},
}

@article{silva_interfaces_2017,
	title = {Interfaces of uniform momentum zones in turbulent boundary layers},
	volume = {820},
	issn = {0022-1120, 1469-7645},
	url = {https://www.cambridge.org/core/journals/journal-of-fluid-mechanics/article/interfaces-of-uniform-momentum-zones-in-turbulent-boundary-layers/49FAE8E736B827592C78F81704AF8CF0},
	doi = {10.1017/jfm.2017.197},
	language = {en},
	urldate = {2026-07-21},
	journal = {Journal of Fluid Mechanics},
	author = {Silva, Charitha M. de and Philip, Jimmy and Hutchins, Nicholas and Marusic, Ivan},
	month = jun,
	year = {2017},
	pages = {451--478},
}

@article{silva_uniform_2016,
	title = {Uniform momentum zones in turbulent boundary layers},
	volume = {786},
	issn = {0022-1120, 1469-7645},
	url = {https://www.cambridge.org/core/journals/journal-of-fluid-mechanics/article/uniform-momentum-zones-in-turbulent-boundary-layers/3656018BE8E3EFCC3FE01245D79ED7D2},
	doi = {10.1017/jfm.2015.672},
	language = {en},
	urldate = {2026-07-21},
	journal = {Journal of Fluid Mechanics},
	author = {Silva, Charitha M. de and Hutchins, Nicholas and Marusic, Ivan},
	month = jan,
	year = {2016},
	pages = {309--331},
}

@article{meinhart_existence_1995,
	title = {On the existence of uniform momentum zones in a turbulent boundary layer},
	volume = {7},
	issn = {1070-6631},
	url = {https://doi.org/10.1063/1.868594},
	doi = {10.1063/1.868594},
	number = {4},
	urldate = {2026-07-21},
	journal = {Physics of Fluids},
	author = {Meinhart, Carl D. and Adrian, Ronald J.},
	month = apr,
	year = {1995},
	pages = {694--696},
}

@article{hoyas_deep-learning-based_2025,
	title = {Deep-learning-based assessment of skin friction in wall-bounded turbulence},
	volume = {10},
	url = {https://link.aps.org/doi/10.1103/b36b-m5hd},
	doi = {10.1103/b36b-m5hd},
	number = {6},
	urldate = {2026-07-21},
	journal = {Physical Review Fluids},
	publisher = {American Physical Society},
	author = {Hoyas, Sergio and Benedikt, Nils and Cremades, Andres and Vinuesa, Ricardo},
	month = jun,
	year = {2025},
	pages = {L062601},
}

@article{houra_effects_2000,
	title = {Effects of adverse pressure gradient on quasi-coherent structures in turbulent boundary layer},
	volume = {21},
	issn = {0142-727X},
	url = {https://www.sciencedirect.com/science/article/pii/S0142727X0000014X},
	doi = {10.1016/S0142-727X(00)00014-X},
	number = {3},
	urldate = {2026-07-21},
	journal = {International Journal of Heat and Fluid Flow},
	author = {Houra, Tomoya and Tsuji, Toshihiro and Nagano, Yasutaka},
	month = jun,
	year = {2000},
	pages = {304--311},
}

@article{hamilton_regeneration_1995,
	title = {Regeneration mechanisms of near-wall turbulence structures},
	volume = {287},
	issn = {1469-7645, 0022-1120},
	url = {https://www.cambridge.org/core/journals/journal-of-fluid-mechanics/article/abs/regeneration-mechanisms-of-nearwall-turbulence-structures/3A4EDE208C75EFC5250278B4D11DC0F9},
	doi = {10.1017/S0022112095000978},
	language = {en},
	urldate = {2026-06-10},
	journal = {Journal of Fluid Mechanics},
	author = {Hamilton, James M. and Kim, John and Waleffe, Fabian},
	month = mar,
	year = {1995},
	pages = {317--348},
}

@article{waleffe_self-sustaining_1997,
	title = {On a self-sustaining process in shear flows},
	volume = {9},
	issn = {1070-6631},
	url = {https://pubs.aip.org/aip/pof/article/9/4/883/260272/On-a-self-sustaining-process-in-shear-flows},
	doi = {10.1063/1.869185},
	language = {en},
	number = {4},
	urldate = {2026-06-10},
	journal = {Physics of Fluids},
	publisher = {AIP Publishing},
	author = {Waleffe, Fabian},
	month = apr,
	year = {1997},
	pages = {883--900},
}

@unpublished{beneitez_improving_2025,
	title = {Improving turbulence control through explainable deep learning},
	url = {http://arxiv.org/abs/2504.02354},
	doi = {10.48550/arXiv.2504.02354},
	urldate = {2026-04-30},
	publisher = {arXiv},
	author = {Beneitez, Miguel and Cremades, Andres and Guastoni, Luca and Vinuesa, Ricardo},
	month = apr,
	year = {2025},
	note = {arXiv:2504.02354 [physics]},
}

@article{hoyas_role_2026,
	title = {The role of turbulence in climate mitigation and sustainable engineering},
	volume = {30},
	issn = {2590-1230},
	url = {https://www.sciencedirect.com/science/article/pii/S2590123026017007},
	doi = {10.1016/j.rineng.2026.110666},
	urldate = {2026-04-30},
	journal = {Results in Engineering},
	author = {Hoyas, Sergio and Vinuesa, Ricardo},
	month = jun,
	year = {2026},
	pages = {110666},
}

@unpublished{fischer_nek5000_2008,
	title = {Nek5000: {Open} source spectral element {CFD} solver},
	url = {https://nek5000.mcs.anl.gov/},
	urldate = {2025-05-19},
	author = {Fischer, P. and Kruse, J. and Mullen, J. and Tuffo, H. and Lottes, J. and Kerkemeier, S.},
	year = {2008},
	note = {Software, Argonne National Laboratory},
}

@inproceedings{lecun_handwritten_1989,
	address = {Denver, CO, USA},
	series = {1989},
	title = {Handwritten {Digit} {Recognition} with a {Back}-{Propagation} {Network}},
	volume = {2},
	url = {https://proceedings.neurips.cc/paper_files/paper/1989/hash/53c3bce66e43be4f209556518c2fcb54-Abstract.html},
	urldate = {2025-02-11},
	booktitle = {Advances in {Neural} {Information} {Processing} {Systems}},
	publisher = {Morgan-Kaufmann},
	author = {LeCun, Yann and Boser, Bernhard and Denker, John and Henderson, Donnie and Howard, R. and Hubbard, Wayne and Jackel, Lawrence},
	month = nov,
	year = {1989},
	pages = {396--404},
}

@article{cremades_classically_2025,
	title = {Classically studied coherent structures only paint a partial picture of wall-bounded turbulence},
	volume = {16},
	copyright = {2025 The Author(s)},
	issn = {2041-1723},
	url = {https://www.nature.com/articles/s41467-025-65199-9},
	doi = {10.1038/s41467-025-65199-9},
	language = {en},
	number = {1},
	urldate = {2025-11-28},
	journal = {Nature Communications},
	author = {Cremades, Andrés and Hoyas, Sergio and Vinuesa, Ricardo},
	month = nov,
	year = {2025},
	note = {Publisher: Nature Publishing Group},
	pages = {10189},
}

@inproceedings{lundberg_unified_2017,
	title = {A {Unified} {Approach} to {Interpreting} {Model} {Predictions}},
	volume = {30},
	url = {https://papers.nips.cc/paper_files/paper/2017/hash/8a20a8621978632d76c43dfd28b67767-Abstract.html},
	urldate = {2023-11-23},
	booktitle = {Advances in {Neural} {Information} {Processing} {Systems}},
	publisher = {Curran Associates, Inc.},
	author = {Lundberg, S. and Lee, S.I.},
	year = {2017},
	pages = {4765--4774},
}

@article{tanarro_effect_2020,
	title = {Effect of adverse pressure gradients on turbulent wing boundary layers},
	volume = {883},
	issn = {0022-1120, 1469-7645},
	url = {https://www.cambridge.org/core/journals/journal-of-fluid-mechanics/article/effect-of-adverse-pressure-gradients-on-turbulent-wing-boundary-layers/47B45FF5F6A4521B826E6D27B1486584},
	doi = {10.1017/jfm.2019.838},
	language = {en},
	urldate = {2025-07-30},
	journal = {Journal of Fluid Mechanics},
	author = {Tanarro, A and Vinuesa, R. and Schlatter, P.},
	month = jan,
	year = {2020},
	pages = {A8},
}

@article{skare_turbulent_1994,
	title = {A {Turbulent} {Equilibrium} {Boundary} {Layer} {Near} {Separation}},
	volume = {272},
	issn = {0022-1120},
	doi = {10.1017/S0022112094004489},
	language = {English},
	journal = {Journal of Fluid Mechanics},
	author = {Skare, Per Egil and Krogstad, Per-Age},
	year = {1994},
	pages = {319--348},
}

@article{alfredsson_new_2011,
	title = {A new scaling for the streamwise turbulence intensity in wall-bounded turbulent flows and what it tells us about the “outer” peak},
	volume = {23},
	issn = {1070-6631},
	url = {https://doi.org/10.1063/1.3581074},
	doi = {10.1063/1.3581074},
	number = {4},
	urldate = {2025-08-06},
	journal = {Physics of Fluids},
	author = {Alfredsson, P. Henrik and Segalini, Antonio and Örlü, Ramis},
	month = apr,
	year = {2011},
	pages = {041702},
}

@article{atzori_coherent_2020,
	title = {Coherent structures in turbulent boundary layers over an airfoil},
	volume = {1522},
	issn = {1742-6596},
	url = {https://dx.doi.org/10.1088/1742-6596/1522/1/012020},
	doi = {10.1088/1742-6596/1522/1/012020},
	language = {en},
	number = {1},
	urldate = {2025-07-24},
	journal = {Journal of Physics: Conference Series},
	author = {Atzori, Marco and Vinuesa, Ricardo and Lozano-Durán, Adrián and Schlatter, Philipp},
	month = apr,
	year = {2020},
	note = {Publisher: IOP Publishing},
	pages = {012020},
}

@article{monty_parametric_2011,
	series = {8th {International} {Symposium} on {Engineering} {Turbulence} {Modelling} and {Measurements}, {Marseille}, {France},{June} 9 to 11, 2010},
	title = {A parametric study of adverse pressure gradient turbulent boundary layers},
	volume = {32},
	issn = {0142-727X},
	url = {https://www.sciencedirect.com/science/article/pii/S0142727X11000452},
	doi = {10.1016/j.ijheatfluidflow.2011.03.004},
	number = {3},
	urldate = {2025-08-01},
	journal = {International Journal of Heat and Fluid Flow},
	author = {Monty, J. P. and Harun, Z. and Marusic, I.},
	month = jun,
	year = {2011},
	pages = {575--585},
}

@article{spalart_experimental_1993,
	title = {Experimental and numerical study of a turbulent boundary layer with pressure gradients},
	volume = {249},
	issn = {1469-7645, 0022-1120},
	url = {https://www.cambridge.org/core/journals/journal-of-fluid-mechanics/article/abs/experimental-and-numerical-study-of-a-turbulent-boundary-layer-with-pressure-gradients/4F085B0AD6B1EA452803FFACA83A389A},
	doi = {10.1017/S002211209300120X},
	language = {en},
	urldate = {2025-08-01},
	journal = {Journal of Fluid Mechanics},
	author = {Spalart, Philippe R. and Watmuff, Jonathan H.},
	month = apr,
	year = {1993},
	pages = {337--371},
}

@article{maciel_coherent_2017,
	title = {Coherent {Structures} in a {Non}-equilibrium {Large}-{Velocity}-{Defect} {Turbulent} {Boundary} {Layer}},
	volume = {98},
	issn = {1573-1987},
	url = {https://doi.org/10.1007/s10494-016-9737-2},
	doi = {10.1007/s10494-016-9737-2},
	language = {en},
	number = {1},
	urldate = {2025-08-01},
	journal = {Flow, Turbulence and Combustion},
	author = {Maciel, Yvan and Simens, Mark P. and Gungor, Ayse G.},
	month = jan,
	year = {2017},
	pages = {1--20},
}

@article{atzori_control_2022,
	title = {Control effects on coherent structures in a non-uniform adverse-pressure-gradient boundary layer},
	volume = {97},
	issn = {0142-727X},
	url = {https://www.sciencedirect.com/science/article/pii/S0142727X22001060},
	doi = {10.1016/j.ijheatfluidflow.2022.109036},
	urldate = {2024-12-20},
	journal = {International Journal of Heat and Fluid Flow},
	author = {Atzori, Marco and Vinuesa, Ricardo and Schlatter, Philipp},
	month = oct,
	year = {2022},
	pages = {109036},
}

@article{vinuesa_revisiting_2017,
	title = {Revisiting {History} {Effects} in {Adverse}-{Pressure}-{Gradient} {Turbulent} {Boundary} {Layers}},
	volume = {99},
	issn = {1573-1987},
	doi = {10.1007/s10494-017-9845-7},
	language = {eng},
	number = {3},
	journal = {Flow, Turbulence and Combustion},
	author = {Vinuesa, Ricardo and Örlü, Ramis and Sanmiguel Vila, Carlos and Ianiro, Andrea and Discetti, Stefano and Schlatter, Philipp},
	year = {2017},
	pmid = {30069157},
	pmcid = {PMC6044237},
	pages = {565--587},
}

@article{vinuesa_determining_2016,
	title = {On determining characteristic length scales in pressure-gradient turbulent boundary layers},
	volume = {28},
	issn = {1070-6631},
	url = {https://pubs.aip.org/aip/pof/article/28/5/055101/1079422/On-determining-characteristic-length-scales-in},
	doi = {10.1063/1.4947532},
	language = {en},
	number = {5},
	urldate = {2025-07-31},
	journal = {Physics of Fluids},
	author = {Vinuesa, R. and Bobke, A. and Örlü, R. and Schlatter, P.},
	month = may,
	year = {2016},
	note = {Publisher: AIP Publishing},
}

@article{schlatter_turbulent_2012,
	title = {Turbulent boundary layers at moderate {Reynolds} numbers: inflow length and tripping effects},
	volume = {710},
	issn = {1469-7645, 0022-1120},
	shorttitle = {Turbulent boundary layers at moderate {Reynolds} numbers},
	url = {https://www.cambridge.org/core/journals/journal-of-fluid-mechanics/article/turbulent-boundary-layers-at-moderate-reynolds-numbers-inflow-length-and-tripping-effects/ADA0DABCC2F2AA0C590939A2A53BB556},
	doi = {10.1017/jfm.2012.324},
	language = {en},
	urldate = {2025-05-19},
	journal = {Journal of Fluid Mechanics},
	author = {Schlatter, Philipp and Örlü, Ramis},
	month = nov,
	year = {2012},
	pages = {5--34},
}

@techreport{iata_global_2024,
	address = {Montreal, Quebec, Canada},
	title = {Global {Outlook} for {Air} {Transport} - {Deep} {Change}},
	url = {https://www.iata.org/en/iata-repository/publications/economic-reports/global-outlook-for-air-transport-june-2024-report/},
	institution = {International Air Transport Association},
	author = {{IATA}},
	month = jun,
	year = {2024},
}

@article{erion_improving_2021,
	title = {Improving performance of deep learning models with axiomatic attribution priors and expected gradients},
	volume = {3},
	copyright = {2021 The Author(s), under exclusive licence to Springer Nature Limited},
	issn = {2522-5839},
	url = {https://www.nature.com/articles/s42256-021-00343-w},
	doi = {10.1038/s42256-021-00343-w},
	language = {en},
	number = {7},
	urldate = {2025-02-04},
	journal = {Nature Machine Intelligence},
	publisher = {Nature Publishing Group},
	author = {Erion, Gabriel and Janizek, Joseph D. and Sturmfels, Pascal and Lundberg, Scott M. and Lee, Su-In},
	month = jul,
	year = {2021},
	pages = {620--631},
}

@article{bobke_history_2017,
	title = {History effects and near equilibrium in adverse-pressure-gradient turbulent boundary layers},
	volume = {820},
	issn = {0022-1120, 1469-7645},
	url = {https://www.cambridge.org/core/journals/journal-of-fluid-mechanics/article/history-effects-and-near-equilibrium-in-adversepressuregradient-turbulent-boundary-layers/39C38082C380F396D004B65F438C296A#figures},
	doi = {10.1017/jfm.2017.236},
	language = {en},
	urldate = {2025-07-30},
	journal = {Journal of Fluid Mechanics},
	author = {Bobke, A. and Vinuesa, R. and Örlü, R. and Schlatter, P.},
	month = jun,
	year = {2017},
	pages = {667--692},
}

@article{vinuesa_turbulent_2018,
	title = {Turbulent boundary layers around wing sections up to {Rec}=1,000,000},
	volume = {72},
	issn = {0142-727X},
	url = {https://www.sciencedirect.com/science/article/pii/S0142727X17311426},
	doi = {10.1016/j.ijheatfluidflow.2018.04.017},
	urldate = {2024-12-20},
	journal = {International Journal of Heat and Fluid Flow},
	author = {Vinuesa, R. and Negi, P. S. and Atzori, M. and Hanifi, A. and Henningson, D. S. and Schlatter, P.},
	month = aug,
	year = {2018},
	pages = {86--99},
}

@article{vinuesa_pressure-gradient_2017,
	title = {Pressure-{Gradient} {Turbulent} {Boundary} {Layers} {Developing} {Around} a {Wing} {Section}},
	volume = {99},
	issn = {1573-1987},
	url = {https://doi.org/10.1007/s10494-017-9840-z},
	doi = {10.1007/s10494-017-9840-z},
	language = {en},
	number = {3},
	urldate = {2025-05-21},
	journal = {Flow, Turbulence and Combustion},
	author = {Vinuesa, Ricardo and Hosseini, Seyed M. and Hanifi, Ardeshir and Henningson, Dan S. and Schlatter, Philipp},
	month = dec,
	year = {2017},
	pages = {613--641},
}

@article{molina-casino_inferring_2025,
	title = {Inferring wall-bounded coherent structures from two-dimensional turbulent fields via {SHAP} analysis},
	volume = {114},
	issn = {0997-7546},
	url = {https://www.sciencedirect.com/science/article/pii/S0997754625000858},
	doi = {10.1016/j.euromechflu.2025.204304},
	urldate = {2025-06-13},
	journal = {European Journal of Mechanics - B/Fluids},
	author = {Molina-Casino, Samuel and Cremades, Andrés and Hoyas, Sergio and Cardesa, José I. and Chedevergne, François and Vinuesa, Ricardo},
	month = nov,
	year = {2025},
	pages = {204304},
}

@article{choi_grid-point_2012,
	title = {Grid-point requirements for large eddy simulation: {Chapman}’s estimates revisited},
	volume = {24},
	issn = {1070-6631},
	shorttitle = {Grid-point requirements for large eddy simulation},
	url = {https://doi.org/10.1063/1.3676783},
	doi = {10.1063/1.3676783},
	number = {1},
	urldate = {2025-05-06},
	journal = {Physics of Fluids},
	author = {Choi, Haecheon and Moin, Parviz},
	month = jan,
	year = {2012},
	pages = {011702},
}

@article{hosseini_direct_2016,
	title = {Direct numerical simulation of the flow around a wing section at moderate {Reynolds} number},
	volume = {61},
	issn = {0142727X},
	url = {https://linkinghub.elsevier.com/retrieve/pii/S0142727X16300169},
	doi = {10.1016/j.ijheatfluidflow.2016.02.001},
	language = {en},
	urldate = {2024-12-20},
	journal = {International Journal of Heat and Fluid Flow},
	author = {Hosseini, S.M. and Vinuesa, R. and Schlatter, P. and Hanifi, A. and Henningson, D.S.},
	month = oct,
	year = {2016},
	pages = {117--128},
}

@article{negi_unsteady_2018,
	title = {Unsteady aerodynamic effects in small-amplitude pitch oscillations of an airfoil},
	volume = {71},
	issn = {0142-727X},
	url = {https://www.sciencedirect.com/science/article/pii/S0142727X1731144X},
	doi = {10.1016/j.ijheatfluidflow.2018.04.009},
	urldate = {2025-04-29},
	journal = {International Journal of Heat and Fluid Flow},
	author = {Negi, P. S. and Vinuesa, R. and Hanifi, A. and Schlatter, P. and Henningson, D. S.},
	month = jun,
	year = {2018},
	pages = {378--391},
}

@article{cremades_additive-feature-attribution_2025,
	title = {Additive-feature-attribution methods: {A} review on explainable artificial intelligence for fluid dynamics and heat transfer},
	volume = {112},
	issn = {0142-727X},
	shorttitle = {Additive-feature-attribution methods},
	url = {https://www.sciencedirect.com/science/article/pii/S0142727X24003874},
	doi = {10.1016/j.ijheatfluidflow.2024.109662},
	urldate = {2024-12-17},
	journal = {International Journal of Heat and Fluid Flow},
	author = {Cremades, Andrés and Hoyas, Sergio and Vinuesa, Ricardo},
	month = mar,
	year = {2025},
	pages = {109662},
}

@article{wallace_wall_1972,
	title = {The wall region in turbulent shear flow},
	volume = {54},
	issn = {1469-7645, 0022-1120},
	url = {https://www.cambridge.org/core/journals/journal-of-fluid-mechanics/article/abs/wall-region-in-turbulent-shear-flow/F38FABEC434430051BE92BAEDAE3EC14},
	doi = {10.1017/S0022112072000515},
	language = {en},
	number = {1},
	urldate = {2025-02-05},
	journal = {Journal of Fluid Mechanics},
	author = {Wallace, James M. and Eckelmann, Helmut and Brodkey, Robert S.},
	month = jul,
	year = {1972},
	pages = {39--48},
}

@incollection{shapley_value_1953,
	title = {A {Value} for n-{Person} {Games}},
	isbn = {978-1-4008-8197-0},
	url = {https://www.degruyter.com/document/doi/10.1515/9781400881970-018/html},
	doi = {10.1515/9781400881970-018},
	language = {en},
	urldate = {2025-02-04},
	booktitle = {Contributions to the {Theory} of {Games}, {Volume} {II}},
	publisher = {Princeton University Press},
	author = {Shapley, L. S.},
	editor = {Kuhn, Harold William and Tucker, Albert William},
	year = {1953},
	pages = {307--318},
}

@inproceedings{ronneberger_u-net_2015,
	address = {Cham},
	title = {U-{Net}: {Convolutional} {Networks} for {Biomedical} {Image} {Segmentation}},
	isbn = {978-3-319-24574-4},
	shorttitle = {U-{Net}},
	doi = {10.1007/978-3-319-24574-4\_28},
	language = {en},
	booktitle = {Medical {Image} {Computing} and {Computer}-{Assisted} {Intervention} – {MICCAI} 2015},
	publisher = {Springer International Publishing},
	author = {Ronneberger, O. and Fischer, P. and Brox, T.},
	editor = {Navab, N. and Hornegger, J. and Wells, W. M. and Frangi, A. F.},
	year = {2015},
	pages = {234--241},
}

@article{chong_general_1990,
	title = {A general classification of three‐dimensional flow fields},
	volume = {2},
	issn = {0899-8213},
	url = {https://doi.org/10.1063/1.857730},
	doi = {10.1063/1.857730},
	number = {5},
	urldate = {2025-01-24},
	journal = {Physics of Fluids A: Fluid Dynamics},
	author = {Chong, M. S. and Perry, A. E. and Cantwell, B. J.},
	month = may,
	year = {1990},
	pages = {765--777},
}

@article{kline_structure_1967,
	title = {The structure of turbulent boundary layers},
	volume = {30},
	issn = {1469-7645, 0022-1120},
	url = {https://www.cambridge.org/core/journals/journal-of-fluid-mechanics/article/structure-of-turbulent-boundary-layers/D0DEB24FDF12498D1574F36FA976B688},
	doi = {10.1017/S0022112067001740},
	language = {en},
	number = {4},
	urldate = {2025-01-24},
	journal = {Journal of Fluid Mechanics},
	author = {Kline, S. J. and Reynolds, W. C. and Schraub, F. A. and Runstadler, P. W.},
	month = dec,
	year = {1967},
	pages = {741--773},
}

@article{brandt_lift-up_2014,
	series = {Enok {Palm} {Memorial} {Volume}},
	title = {The lift-up effect: {The} linear mechanism behind transition and turbulence in shear flows},
	volume = {47},
	issn = {0997-7546},
	shorttitle = {The lift-up effect},
	url = {https://www.sciencedirect.com/science/article/pii/S0997754614000405},
	doi = {10.1016/j.euromechflu.2014.03.005},
	urldate = {2025-01-24},
	journal = {European Journal of Mechanics - B/Fluids},
	author = {Brandt, Luca},
	month = sep,
	year = {2014},
	pages = {80--96},
}

@article{jeong_coherent_1997,
	title = {Coherent structures near the wall in a turbulent channel flow},
	volume = {332},
	issn = {0022-1120, 1469-7645},
	url = {https://www.cambridge.org/core/journals/journal-of-fluid-mechanics/article/coherent-structures-near-the-wall-in-a-turbulent-channel-flow/01E20E6B9FE02829717DB720AD3EC416},
	doi = {10.1017/S0022112096003965},
	language = {en},
	urldate = {2025-01-24},
	journal = {Journal of Fluid Mechanics},
	author = {Jeong, J. and Hussain, F. and Schoppa, W. and Kim, J.},
	month = feb,
	year = {1997},
	pages = {185--214},
}

@article{cremades_identifying_2024,
	title = {Identifying regions of importance in wall-bounded turbulence through explainable deep learning},
	volume = {15},
	issn = {2041-1723},
	url = {https://www.nature.com/articles/s41467-024-47954-6},
	doi = {10.1038/s41467-024-47954-6},
	number = {1},
	journal = {Nature Communications},
	author = {Cremades, A. and Hoyas, S. and Deshpande, R. and Quintero, P. and Lellep, M. and Lee, W. J. and Monty, J. P. and Hutchins, N. and Linkmann, M. and Marusic, I. and Vinuesa, R.},
	year = {2024},
	pages = {3864},
}

@article{lozano-duran_three-dimensional_2012,
	title = {The three-dimensional structure of momentum transfer in turbulent channels},
	volume = {694},
	issn = {1469-7645, 0022-1120},
	url = {https://www.cambridge.org/core/journals/journal-of-fluid-mechanics/article/threedimensional-structure-of-momentum-transfer-in-turbulent-channels/320CEAE9CB3FBC95685DF68A8DDA7F31},
	doi = {10.1017/jfm.2011.524},
	language = {en},
	urldate = {2023-11-23},
	journal = {Journal of Fluid Mechanics},
	author = {Lozano-Durán, Adrián and Flores, Oscar and Jiménez, Javier},
	month = mar,
	year = {2012},
	pages = {100--130},
}

@article{jimenez_coherent_2018,
	title = {Coherent structures in wall-bounded turbulence},
	volume = {842},
	issn = {0022-1120, 1469-7645},
	url = {https://www.cambridge.org/core/journals/journal-of-fluid-mechanics/article/coherent-structures-in-wallbounded-turbulence/8B21CB7E6A285152AE07230624BB5F98},
	doi = {10.1017/jfm.2018.144},
	language = {en},
	urldate = {2023-11-23},
	journal = {Journal of Fluid Mechanics},
	author = {Jiménez, Javier},
	month = may,
	year = {2018},
	pages = {P1},
}
